%% file: VLDB-2026-DAGSmith-Dependency-Aware-Rewriting-for-dbt-Style-SQL-Pipelines.tex
\documentclass[sigconf, nonacm]{acmart}

\newcommand\vldbpagestyle{plain}

\input{imports}
\input{macros}

\begin{document}

\raggedbottom

\title{\tool: Dependency-Aware Rewriting for dbt-Style SQL Pipelines}

%%
%% The "author" command and its associated commands are used to define the authors and their affiliations.

\author{Jie Liu}
\email{jiezzliu@umich.edu}
\affiliation{%
  \department{Computer Science and Engineering}
  \institution{University of Michigan, Ann Arbor}
  \country{USA}
}

\author{Lin Ma}
\email{linmacse@umich.edu}
\affiliation{%
  \department{Computer Science and Engineering}
  \institution{University of Michigan, Ann Arbor}
  \country{USA}
}

\author{Barzan Mozafari}
\email{mozafari@umich.edu}
\affiliation{%
  \department{Computer Science and Engineering}
  \institution{University of Michigan, Ann Arbor}
  \country{USA}
}

\renewcommand{\shortauthors}{Liu, Ma, and Mozafari}

\include{sections/0_abstract}

\maketitle

\pagestyle{\vldbpagestyle}

\input{sections/1_intro}

\input{sections/2_motivation}
\input{sections/3_approach}
\input{sections/4_eval}
\input{sections/5_related}

\input{sections/6_conclusion}

% \clearpage

% \input{sections/x_ILP}

\vspace{0.75\baselineskip}
\noindent\textbf{Disclaimer.} This draft was polished with AI assistance for spelling, clarity, and condensation; the authors take full responsibility for the accuracy of the ideas, claims, and technical content.

\bibliographystyle{ACM-Reference-Format}
\bibliography{main}

\end{document}

%% file: imports.tex
\usepackage{amsmath,amssymb,mathtools}
\usepackage{algorithm}
\usepackage[noend]{algpseudocode}

\usepackage{soul}
\usepackage{enumitem}
\usepackage{multirow} 
\usepackage{makecell}
\usepackage{subcaption}
\usepackage[page]{appendix}
\usepackage{graphicx}  % for \resizebox
\usepackage{hyperref}
\usepackage[nameinlink,capitalise]{cleveref}
\usepackage{listings}
\usepackage{framed}
\usepackage{xspace}
\usepackage{tikz}
\usetikzlibrary{arrows.meta, positioning, shapes.geometric, fit, calc, backgrounds}
\usepackage{xcolor}
\usepackage[normalem]{ulem}

%% file: macros.tex
\newcommand{\head}[1]{{{\vspace*{1mm} \noindent \textbf{#1.}}}}
\newcommand{\ph}[1]{\vspace{1mm} \noindent \textbf{#1} ---}

\newcommand{\RememberToChange}[1]{{#1}}
\newcommand{\code}[1]{{\normalfont\ttfamily\small #1}}

\newcommand{\tool}{{DAGSmith}\xspace}

\newfloat{listing}{htbp}{lol}
\floatname{listing}{Listing}

\newcommand{\figsubheader}[1]{%
    {\centering\small\textbf{#1}\par\medskip}%
}

\definecolor{lin}{HTML}{e31a1c}
\newcommand{\linc}[1]{\textcolor{lin}{lin: #1}}

\newcommand{\jie}[1]{\textcolor{blue}{Jie: #1}}

\newcommand{\inv}{\vspace*{-2mm}}

%% file: sections/0_abstract.tex
% !TEX root = ../VLDB-2026-DAGSmith-Dependency-Aware-Rewriting-for-dbt-Style-SQL-Pipelines.tex
\begin{abstract}
Modern analytics is increasingly organized as recurring SQL pipelines rather than isolated SQL statements. Tools such  as dbt---which have gained extreme popularity in recent years---allow teams to write each transformation as SQL and make dependencies between transformations explicit, producing directed acyclic graphs (DAGs) with hundreds or thousands of interdependent SQL models. Existing techniques are not effective at optimizing these expensive pipelines. Traditional query optimizers and source-to-source query rewriters operate on one query at a time, while materialized-view selection and multi-query optimization address narrower forms of reuse. They do not directly exploit the pipeline-level information exposed by explicit dependencies: how intermediate results are consumed, which downstream outputs depend on each computation, where expensive work occurs relative to data reduction, which intermediate results are worth persisting, and how refresh schedules relate to the rate at which inputs change and outputs are consumed.

We introduce \tool, to the best of our knowledge the first holistic dependency-aware source-to-source rewriting system for SQL pipeline DAGs. \tool treats explicit dependencies as optimization signals.
It analyzes each transformation together with its upstream inputs, downstream consumers, and position in the pipeline DAG, uses an LLM to propose pipeline-level refactorings, separates SQL generation and data-backed equivalence checks to reject unsafe rewrites, retunes persistence choices with a learned cost model, and 
selects a globally compatible set of rewrites that avoids conflicts across transformations.
This enables dependency-edge simplification, non-local semantic reuse, downstream-aware pruning, pipeline-aware work placement, rewrite-materialization co-optimization, and frequency-aware optimization while keeping the pipeline executable by the same SQL engine and orchestration framework. Our extensive evaluation shows that, on the open-source Tuva dbt project, \tool reduces elapsed time by \RememberToChange{42.6\%} and warehouse compute cost by \RememberToChange{67.7\%}; these reductions are \RememberToChange{22.8\%}/\RememberToChange{83.0\%} larger than materialization-only tuning and \RememberToChange{98.1\%}/\RememberToChange{348.3\%} larger than state-of-the-art single-query rewriting techniques.
\end{abstract}

%% file: sections/1_intro.tex
% !TEX root = ../VLDB-2026-DAGSmith-Dependency-Aware-Rewriting-for-dbt-Style-SQL-Pipelines.tex
\section{Introduction}\label{sec:intro}

\ph{Rising SQL Complexity}
SQL is no longer only a hand-written interface for individual analytical questions. It is increasingly the implementation language for recurring data products: governed dashboards, AI-ready feature tables, self-service BI datasets, and organization-wide reporting layers. At the same time, more SQL is produced by BI systems, code generators, and AI-assisted development tools rather than written from scratch.  In fact, recent studies report  that 70\% of analytics professionals use AI for code development~\cite{dbt2025state}, and that  BI-generated SQL  contain extremely complex scalar expressions and relational operator trees~\cite{getreal2018}.
This has created a major efficiency gap between the complexity of the SQL queries generated and what today's query optimizers are capable of effectively optimizing. Source-to-source query rewriting is the traditional way to narrow this gap: a rewriter transforms complex SQL into an equivalent form that the query optimizer is more likely to turn into an efficient plan. However, query rewriting has traditionally treated one SQL statement as the unit of optimization. Unfortunately, the effectiveness of this one-query-at-a-time approach is increasingly limited by a second trend: modern SQL is no longer produced and executed only as isolated statements, but as recurring transformation pipelines.

\begin{figure}[t]
\centering
\begin{tikzpicture}[
  font=\scriptsize,
  node distance=5mm and 4mm,
  box/.style={draw=black!70, text=black, rounded corners=2pt, align=left,
              inner sep=2.2pt, minimum width=22mm},
  edge/.style={-{Stealth[length=1.5mm]}, black!70, thin}
]
\node[box, minimum width=17mm] (orders) {\textbf{orders}\\raw table};
\node[box, minimum width=24mm, right=of orders] (clean) {\textbf{clean\_orders}\\\code{SELECT ...}\\\code{FROM orders}\\\code{WHERE status='paid'}};
\node[box, minimum width=20mm, above right=of clean] (rev) {\textbf{daily\_revenue}\\\code{GROUP BY day}};
\node[box, minimum width=20mm, below right=of clean] (cust) {\textbf{active\_customers}\\\code{SELECT DISTINCT}\\\code{customer\_id}};
\draw[edge] (orders) -- (clean);
\draw[edge] (clean) -- (rev);
\draw[edge] (clean) -- (cust);
\end{tikzpicture}
% \inv
\caption{A toy SQL pipeline DAG. Nodes are SQL transformations or SQL-produced tables (called ``models'' in dbt); edges show which transformation consumes another transformation's output.}
\label{fig:intro-toy}
\inv
\end{figure}
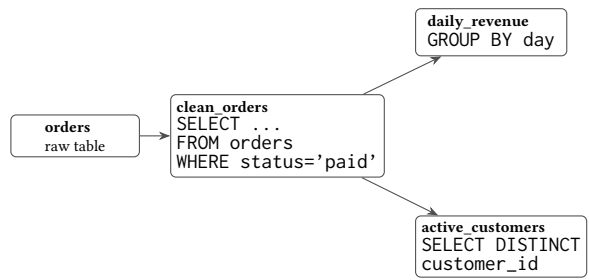

\ph{Rise of SQL Pipelines}
Data teams are increasingly building large, recurring SQL pipelines. This growth is driven by complex business logic that must be computed repeatedly and reused consistently across governed dashboards, AI-ready feature tables, self-service BI datasets, compliance reports, and downstream applications.
 One of the most widely adopted tools for these workflows is \emph{dbt} (the data build tool)~\cite{dbtWhatExactly}, reported to be used by 50K+ data teams weekly across the world~\cite{dbtComplexity}. 
dbt lets teams write transformations as version-controlled SQL \code{SELECT} statements (called ``models'' in dbt, which we use interchangeably in this paper), declare dependencies among them, and compile the resulting graph into SQL jobs executed by a data warehouse.  Figure~\ref{fig:intro-toy} gives a toy example. A raw orders table feeds a cleaning transformation; the cleaned result feeds both a revenue table and a daily active-customer table. The SQL text explains each local transformation, while the edges explain how the transformations compose into a pipeline.
This trend is much broader than dbt. SQLMesh~\cite{sqlmeshOverview}, Dagster asset graphs~\cite{dagsterAssets}, Airflow directed acyclic graphs~\cite{airflowDags}, Prefect flows~\cite{prefectTasks}, Argo Workflows~\cite{argoDag}, Databricks Workflows~\cite{databricksJobs}, and medallion lakehouse pipelines~\cite{databricksMedallion} all represent recurring data transformations through explicit dependencies. In this paper, we target the SQL subset of this broader pattern, which we refer to as \emph{SQL pipeline DAG}: a recurring directed acyclic graph whose nodes are SQL transformations or SQL-produced tables/views, and whose edges record that one node reads another node's output.

\ph{Pipeline Optimization Gap} 
These pipeline DAGs often include hundreds or thousands of SQL transformations that reference one another. Their scale makes them difficult to optimize with techniques designed for a single SQL statement or a flat workload: inlining every referenced transformation into each downstream query would produce deeply nested SQL with tens or hundreds of layers whose size and complexity are well beyond what today's optimizers can even tackle. In particular, existing \emph{query rewriting} techniques, whether traditional rule-based systems~\cite{wetune} or modern synthesis- or LLM-based techniques~\cite{slabcity,genrewrite,rbot}, rewrite one SQL statement at a time.  This scope is useful for improving a standalone query, but it is too narrow for optimizations whose correctness or benefit depends on the surrounding pipeline, such as removing work no final output uses, reusing equivalent logic across separate transformations, or changing refresh frequency based on input changes and output consumption.
 \emph{Multi-query optimization} techniques~\cite{sellis1988multiple,roy2000efficient} consider multiple queries together, but their classical goal is to share common scans, joins, or intermediate results across a batch of submitted queries. They do not generally rewrite a SQL pipeline by adding, removing, splitting, or merging transformations based on downstream demand and refresh behavior.
 \emph{Materialized-view selection}~\cite{harinarayan1996implementing,agrawal2000automated} chooses intermediate query results to precompute and rewrites future queries to read those precomputed results. This improves reuse when the relevant computation has already been identified, but it does not by itself determine that a pipeline should drop unused work, factor repeated logic written in different forms, move expensive operations after data reduction, or refresh different parts of the pipeline at different rates.
Lastly, \emph{Workflow schedulers}~\cite{airflowDags,dagsterAssets,prefectTasks,argoDag} know dependencies, but usually optimize execution order, failure recovery, retries, or resource allocation rather than the SQL semantics of the transformations themselves.

\ph{Dependencies as Signals} 
\label{sec:dependencies-as-signals}
Our key insight is that explicit dependencies are not merely execution constraints; they are  signals that can be leveraged for optimization.
In particular, a SQL pipeline DAG exposes six kinds of information that are hidden or much harder to recover when queries are optimized independently: \textbf{(1) Edge semantics:} producer-consumer edges reveal how one SQL transformation's output is used by later transformations; \textbf{(2) Cross-graph redundancy:} similar business logic can appear in distant parts of the graph, even when SQL text differs; \textbf{(3) Downstream demand:} downstream dependencies reveal which columns, joins, filters, and intermediate results can affect final outputs; \textbf{(4) Operator placement:} the graph reveals where expensive operations occur relative to data-reducing filters, projections, joins, and aggregations; 
\textbf{(5) Rewrite-persistence coupling:} unlike single-query rewriting, which is constrained to equivalent rewrites that are cheaper in isolation, pipeline rewriting can consider alternatives that produce additional reusable outputs and are therefore locally more expensive, as long as those outputs are materialized once and reused by enough downstream consumers; and \textbf{(6) Refresh patterns:} schedules and source update rates reveal how often inputs change and outputs are consumed.

\ph{Our Approach}
In this paper, we introduce \tool, the first (to the best of our knowledge) holistic, dependency-aware rewriting system for SQL pipeline DAGs.  \tool exploits the exposed dependency structure to  perform six classes of optimization: \textbf{(1) dependency-edge simplification:} remove or simplify dependency edges whose purpose can be expressed more directly, while preserving final outputs; \textbf{(2) non-local semantic reuse:} factor repeated work into a shared computation across non-adjacent transformations; \textbf{(3) downstream-aware pruning:} remove work that is provably irrelevant to every final output; \textbf{(4) pipeline-aware work placement:} move expensive work after data-reducing steps as long as final outputs are not impacted; \textbf{(5) rewrite-materialization co-optimization:} decide when rewrite benefits are outweighed by materialization cost, and when materializing a rewritten result makes the rewrite worthwhile; and \textbf{(6) frequency-aware optimization:} align computation frequency with input-change rates and output-consumption rates, avoiding refreshes that do not impact consumers.

\tool first analyzes compiled SQL and dependency edges to identify regions likely to contain these opportunities. It then leverages Large Language Models (LLMs) to reason over the relevant subgraph and propose concrete refactorings, but separates proposal, criticism, SQL generation, and data-backed equivalence checking so unsafe or counterproductive rewrites can be rejected before adoption. Adopted candidates are not evaluated in isolation: \tool retunes persistence choices for each candidate using a learned cost model and an iterative local linearization procedure, then solves a global selection problem to choose a non-conflicting subset of rewrites. Finally, frequency information from source update rates and output demand reweights cost estimates, detects over-scheduled work, and generates splitting candidates when slow-changing logic is embedded in fast-refreshing transformations.
The result is a rewritten, output-equivalent SQL pipeline that remains executable by the same SQL engine and orchestration framework, but is significantly more performant.

\ph{Contributions}
This paper makes the following contributions:
\begin{enumerate}[leftmargin=*, topsep=2pt, itemsep=1pt]
    \item We formulate dependency-aware source-to-source rewriting for SQL pipeline DAGs. We present the first holistic approach to this problem (to the best of our knowledge) by identifying the pipeline-level signals exposed by explicit dependencies and the rewriting opportunities they enable: dependency-edge simplification, non-local semantic reuse, downstream-aware pruning, pipeline-aware work placement, rewrite-materialization co-optimization, and frequency-aware optimization.
    \item We present the design of \tool, which combines structural graph analysis, LLM-guided refactoring, separated proposal, criticism, SQL generation, and data-backed equivalence checking, learned cost modeling, persistence tuning, frequency-aware scheduling, and global candidate selection.
    \item We conduct a comprehensive evaluation showing that, on the open-source Tuva dbt project, \tool reduces elapsed time by \RememberToChange{42.6\%} and warehouse compute cost by \RememberToChange{67.7\%}; these reductions are \RememberToChange{22.8\%}/\RememberToChange{83.0\%} larger than materialization-only tuning and \RememberToChange{98.1\%}/\RememberToChange{348.3\%} larger than state-of-the-art single-query rewriting techniques.
   \end{enumerate}

The rest of the paper is organized as follows. Section~\ref{sec:motivating} defines the setting and illustrates the opportunity space. Section~\ref{sec:approach} presents the design of \tool. Section~\ref{sec:eval} presents our evaluation, Section~\ref{sec:related} discusses prior work, and Section~\ref{sec:conclusion} concludes.

%% file: sections/2_motivation.tex
% !TEX root = ../VLDB-2026-DAGSmith-Dependency-Aware-Rewriting-for-dbt-Style-SQL-Pipelines.tex
\section{Problem Setting}
\label{sec:motivating}

Before presenting our approach, we first provide the necessary background on SQL pipeline DAGs (\S\ref{sec:dbt-setting}), then present a few motivating examples of pipeline-level optimization opportunities exposed by explicit dependencies (\S\ref{sec:motivating-examples}). Lastly, we explain why local query rewriting is insufficient for this setting (\S\ref{sec:why-different}).

\subsection{Background: SQL Pipeline DAGs}
\label{sec:dbt-setting}

 A \emph{SQL pipeline DAG} is a recurring SQL program whose nodes are named transformations and whose directed edges record producer-consumer relationships among their outputs (see Figure~\ref{fig:intro-toy} for a toy example). The DAG determines evaluation order, while node metadata can specify whether each result is materialized and how frequently it is refreshed. Because dbt calls each named SQL transformation a \emph{model}, we use \emph{model} for dbt-specific examples and \emph{transformation} for the broader setting.

\ph{Target workload}
Although we use dbt in our examples and evaluation, the optimization problem is more general. \tool targets recurring SQL-based DAGs: pipelines in which node logic is available as SQL, edges expose producer-consumer dependencies, and system metadata records persistence and refresh behavior. dbt is a prominent instance of this setting, but similar dependency graphs arise in SQLMesh~\cite{sqlmeshOverview}, Dagster~\cite{dagsterAssets}, Airflow~\cite{airflowDags}, Prefect~\cite{prefectTasks}, Argo~\cite{argoDag}, Databricks Workflows and Lakeflow Jobs~\cite{databricksJobs}, medallion lakehouse pipelines~\cite{databricksMedallion}, and Luigi~\cite{luigiDocs}.

\subsection{Motivating Examples}
\label{sec:motivating-examples}

% In this section, we use a small claims-processing DAG to illustrate how explicit dependencies expose optimization opportunities that are invisible when each query is rewritten in isolation. For space, the examples combine multiple opportunities in one small claims-processing DAG; in practice, each opportunity can appear independently inside much larger DAGs with hundreds or thousands of models. Figure~\ref{fig:strategy-c-graph}(a) shows the original pipeline. 
% Here, we use a simplified healthcare claims-processing pipeline, inspired by the open-source Tuva project~\cite{tuvaProject}.

We illustrate optimization opportunities on a small healthcare claims-processing DAG inspired by the open-source Tuva project~\cite{tuvaProject}; for space it combines several that in practice appear independently inside much larger DAGs with hundreds or thousands of models. Figure~\ref{fig:strategy-c-graph}(a) shows the original pipeline.
Its main input is \code{claims}, a large table of claim-line records that is refreshed frequently. It feeds into 
service-category models such as \code{dialysis}, \code{psychiatric}, \code{equipment}, and \code{nursing}. These models classify claim lines using SQL predicates over claim attributes; Figure~\ref{fig:strategy-c-sql} gives simplified examples. The \code{nursing} model excludes claim lines already categorized as equipment, so it reads \code{equipment} as an exclusion list. The \code{claims\_summary} model joins claim lines with a 
slow-changing lookup table, \code{code\_catalog}, that maps raw claim codes to descriptions, and then applies code-grouping logic over those descriptions. We next describe three examples that exploit these explicit dependencies; when all applied together, they produce the more efficient DAG in Figure~\ref{fig:strategy-c-graph}(b).

\begin{figure*}[t]
\inv
\centering
{ \input{figures/motivation_dag_graph}}
\inv
\caption{Healthcare SQL pipeline DAG used for three motivating examples. The dotted line separates the original DAG (a) from the cumulative rewritten DAG (b). The rewritten DAG introduces reusable \code{service\_flags}, rewrites downstream models (\code{dialysis}, \code{psychiatric}, and \code{nursing}), removes the \code{equipment} dependency from \code{nursing}, and moves slow-changing reference-code grouping out of the fast claims path. }
\label{fig:strategy-c-graph}
\end{figure*}
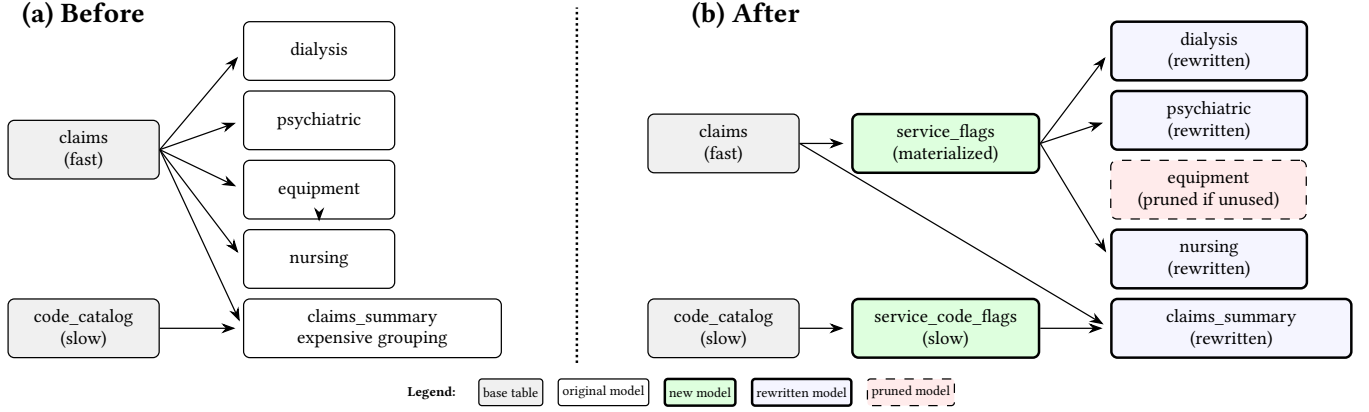

\begin{figure*}[t]
\centering
\captionsetup[subfigure]{justification=centering}
\begin{subfigure}[t]{0.32\textwidth}
\centering
\begin{minipage}[t][51mm][t]{\linewidth}
{\scriptsize\textbf{Exposes:} cross-graph redundancy; reuse context\par}
{\scriptsize\textbf{Enables:} non-local reuse; rewrite-materialization co-optimization\par\smallskip}
{\scriptsize\textbf{Before}\par}
\begin{lstlisting}[language=SQL,basicstyle=\ttfamily\tiny,breaklines=true]
-- dialysis model
SELECT DISTINCT claim_id
FROM claims
WHERE bill_code LIKE '72%'
   OR taxonomy IN (...);
\end{lstlisting}
{\scriptsize\textbf{After}\par}
\begin{lstlisting}[language=SQL,basicstyle=\ttfamily\tiny,breaklines=true]
-- new service_flags model
SELECT claim_id,
  MAX(bill_code LIKE '72%' OR taxonomy IN (...)) AS f_dia,
  MAX(diagnosis BETWEEN '290' AND '319') AS f_psy
FROM claims;

-- rewritten dialysis model
SELECT claim_id FROM service_flags
WHERE f_dia;
\end{lstlisting}
\end{minipage}
\caption{Example 1: shared service flags}
\label{fig:strategy-c-sql-reuse}
\end{subfigure}\hfill
\begin{subfigure}[t]{0.32\textwidth}
\centering
\begin{minipage}[t][51mm][t]{\linewidth}
{\scriptsize\textbf{Exposes:} edge semantics; downstream demand\par}
{\scriptsize\textbf{Enables:} edge simplification; downstream-aware pruning\par\smallskip}
{\scriptsize\textbf{Before}\par}
\begin{lstlisting}[language=SQL,basicstyle=\ttfamily\tiny,breaklines=true]
-- nursing model
SELECT * FROM claims c
WHERE c.facility_code IN ('31','32')
AND NOT EXISTS (
  SELECT 1 FROM equipment e
  WHERE e.claim_id = c.claim_id);
\end{lstlisting}
{\scriptsize\textbf{After}\par}
\begin{lstlisting}[language=SQL,basicstyle=\ttfamily\tiny,breaklines=true]
-- rewritten nursing model
SELECT * FROM claims c
WHERE c.facility_code IN ('31','32')
  AND c.category <> 'equipment';

-- equipment has no remaining consumers
\end{lstlisting}
\end{minipage}
\caption{Example 2: removing an exclusion edge}
\label{fig:strategy-c-sql-edge}
\end{subfigure}\hfill
\begin{subfigure}[t]{0.32\textwidth}
\centering
\begin{minipage}[t][51mm][t]{\linewidth}
{\scriptsize\textbf{Exposes:} refresh patterns; operator placement\par}
{\scriptsize\textbf{Enables:} frequency-aware optimization; work placement\par\smallskip}
{\scriptsize\textbf{Before}\par}
\begin{lstlisting}[language=SQL,basicstyle=\ttfamily\tiny,breaklines=true]
-- claims_summary model
SELECT c.id, group_code(r.description) AS grp
FROM claims c JOIN code_catalog r USING(code);
\end{lstlisting}
{\scriptsize\textbf{After}\par}
\begin{lstlisting}[language=SQL,basicstyle=\ttfamily\tiny,breaklines=true]
-- new service_code_flags model
SELECT code, group_code(description) AS grp
FROM code_catalog;

-- rewritten claims_summary model
SELECT c.id, f.grp
FROM claims c JOIN service_code_flags f USING(code);
\end{lstlisting}
\end{minipage}
\caption{Example 3: precomputing slow-side logic}
\label{fig:strategy-c-sql-frequency}
\end{subfigure}
\caption{SQL fragments for the healthcare examples in Figure~\ref{fig:strategy-c-graph}. Each panel corresponds to one motivating example: (a) Example 1 factors repeated service-category logic into shared flags, (b) Example 2 removes an exclusion-list dependency, and (c) Example 3 moves slow-side grouping out of the fast claims path. Each panel separates the original SQL from the rewritten SQL.}
\label{fig:strategy-c-sql}
\inv
\end{figure*}

\head{Example 1: Cross-graph redundancy + rewrite-persistence coupling}
In Figure~\ref{fig:strategy-c-graph}(a), the fanout from \code{claims} to the service-category models (\code{dialysis}, \code{psychiatric}, \code{equipment}, and \code{nursing})
exposes cross-graph redundancy: several models scan the same large source (i.e., \code{claims}), and evaluate related billing-code, taxonomy-code, or diagnosis-code predicates in separate SQL files. By exploiting this redundancy, one can create the shared \code{service\_flags} model in Figure~\ref{fig:strategy-c-graph}(b) and rewrite each of the downstream models  to read the corresponding flag. Figure~\ref{fig:strategy-c-sql-reuse} shows the rewrite for \code{dialysis}; the other service-category models can be rewritten similarly by reading their own flags.
In \S\ref{sec:dependencies-as-signals}, we call this non-local semantic reuse.

This shared model also creates a materialization choice: computing extra flags such as \code{f\_psy} pays off only when \code{service\_flags} is persisted once and reused, rather than inlined under each consumer. This is rewrite-materialization co-optimization in \S\ref{sec:dependencies-as-signals}.

\head{Example 2: Edge semantics + downstream demand}
We can also exploit the edge semantics in Figure~\ref{fig:strategy-c-graph}(a): the edge from \code{equipment} to \code{nursing} implies that \code{nursing} consumes \code{equipment}, but the SQL definitions show that \code{nursing} uses \code{equipment} only to exclude claim lines whose category is equipment. 
Figure~\ref{fig:strategy-c-sql-edge} shows that the original \code{nursing} SQL anti-joins against \code{equipment}, while the rewrite uses a direct predicate over \code{claims}. This avoids building or scanning \code{equipment} as an exclusion list and removes the anti-join from \code{nursing};  we called this dependency-edge simplification in \S\ref{sec:dependencies-as-signals}.

The graph also exposes downstream demand: after this rewrite, if no remaining edge leaves \code{equipment}, no final output can observe it and the model can be skipped entirely. We referred to this as downstream-aware pruning in \S\ref{sec:dependencies-as-signals}.

\head{Example 3: Refresh + placement}
The \code{claims\_summary} model in Figure~\ref{fig:strategy-c-graph}(a) combines fast-changing claims with slow-changing reference codes and then applies code-grouping logic after the join. 
As Figure~\ref{fig:strategy-c-sql-frequency} shows, \code{group\_code} depends only on the slow-changing \code{code\_catalog}, but the original \code{claims\_summary} evaluates it after joining into the high-volume \code{claims} stream. The recurring DAG exposes source update rates and consumer refresh schedules, so moving \code{group\_code} into \code{service\_code\_flags} lets the grouped lookup be recomputed only when the lookup table changes or consumers need it, while also applying the grouping before the join expands data volume. These are called frequency-aware optimization and pipeline-aware work placement in \S\ref{sec:dependencies-as-signals}.

\subsection{Why Existing Techniques Fall Short}
\label{sec:why-different}

The examples above point to a broader limitation: pipeline rewriting needs information that is distributed across the dependency graph, not only inside the query being rewritten.
Single-query rewrite systems~\cite{wetune,rbot} operate within the boundary of one query and cannot see the upstream queries that define its inputs or the downstream queries that consume its output.
Multi-query optimization and common subexpression elimination~\cite{sellis1988multiple} share intermediate computations across queries but rely on syntactic matching, so they miss the cross-query semantic equivalence illustrated in Figure~\ref{fig:strategy-c-sql-reuse}.
Materialized view selection~\cite{agrawal2000automated} can choose which existing results to persist, but it does not rewrite the DAG to move slow-changing work out of a fast-refresh path, as illustrated in Figure~\ref{fig:strategy-c-sql-frequency}.

% The gap is not that these techniques are poorly designed; it is that none treats an explicit SQL dependency graph as the unit of rewriting.
% Dependency-aware rewriting must combine capabilities that are usually considered separately:
% use producer and consumer context, find semantic reuse beyond syntactic overlap, prune work that no downstream output consumes, move expensive work after data-reducing steps, evaluate rewrites together with persistence choices, and account for recurring refresh behavior.
% The remainder of this paper presents an approach that integrates these capabilities into a coherent rewriting framework.
The gap is not that these techniques are poorly designed; it is that none treats an explicit SQL dependency graph as the unit of rewriting, combining capabilities usually considered separately. The remainder of this paper presents an approach that integrates them into a coherent rewriting framework.

%% file: figures/motivation_dag_graph.tex
\resizebox{\textwidth}{!}{%
\begin{tikzpicture}[
  font=\scriptsize,
  n/.style={draw, rounded corners=2pt, align=center, minimum height=6.6mm, minimum width=17mm, inner sep=1.6pt},
  src/.style={n, fill=gray!12},
  new/.style={n, fill=green!13, thick, minimum width=21mm},
  rew/.style={n, fill=blue!4, thick, minimum width=19mm},
  gone/.style={n, fill=red!8, dashed, minimum width=20mm},
  wide/.style={n, minimum width=29mm},
  legend/.style={draw, rounded corners=1pt, align=center, font=\fontsize{4.4}{5.0}\selectfont, minimum height=3.0mm, inner sep=1.4pt},
  legnew/.style={legend, fill=green!13, thick},
  legrew/.style={legend, fill=blue!4, thick},
  arr/.style={-{Stealth[length=1.6mm,width=1.15mm]}, thin, shorten >=2.4pt}
]
\draw[densely dotted, thick] (5.55,-1.95) -- (5.55,2.05);
% ---------- (a) Before ----------
\node[font=\bfseries] at (0,1.96) {(a) Before};
\node[src] (bc) at (0,0.45) {claims\\(fast)};
\node[src] (bcode) at (0,-1.56) {code\_catalog\\(slow)};
\node[n] (bd) at (2.65,1.56) {dialysis};
\node[n] (bp) at (2.65,0.78) {psychiatric};
\node[n] (be) at (2.65,0.00) {equipment};
\node[n] (bn) at (2.65,-0.78) {nursing};
\node[wide] (bs) at (3.25,-1.56) {claims\_summary\\expensive grouping};
\draw[arr] (bc.east) -- (bd.west);
\draw[arr] (bc.east) -- (bp.west);
\draw[arr] (bc.east) -- (be.west);
\draw[arr] (bc.east) -- (bn.west);
\draw[arr] (be.south) -- (bn.north);
\draw[arr] (bc.east) -- (bs.west);
\draw[arr] (bcode.east) -- (bs.west);
% ---------- (b) After ----------
\node[font=\bfseries] at (7.45,1.96) {(b) After};
\node[src] (ac) at (7.20,0.52) {claims\\(fast)};
\node[src] (acode) at (7.20,-1.56) {code\_catalog\\(slow)};
\node[new] (asf) at (9.70,0.52) {service\_flags\\(materialized)};
\node[new] (asc) at (9.70,-1.56) {service\_code\_flags\\(slow)};
\node[rew, minimum width=22mm, anchor=west] (ad) at (11.55,1.56) {dialysis\\(rewritten)};
\node[rew, minimum width=22mm, anchor=west] (ap) at (11.55,0.78) {psychiatric\\(rewritten)};
\node[rew, minimum width=22mm, anchor=west] (an) at (11.55,-0.78) {nursing\\(rewritten)};
\node[rew, minimum width=27mm, anchor=west] (asum) at (11.55,-1.56) {claims\_summary\\(rewritten)};
\node[gone, minimum width=22mm, anchor=west] (ae) at (11.55,0.00) {equipment\\(pruned if unused)};
\draw[arr] (ac.east) -- (asf.west);
\draw[arr] (asf.east) -- (ad.west);
\draw[arr] (asf.east) -- (ap.west);
\draw[arr] (asf.east) -- (an.west);
\draw[arr] (acode.east) -- (asc.west);
\draw[arr] (ac.east) -- (asum.west);
\draw[arr] (asc.east) -- (asum.west);
% ---------- Legend (single tight row, chained so boxes cannot overlap) ----------
\node[font=\fontsize{4.4}{5.0}\selectfont\bfseries, anchor=east] (leglbl) at (4.30,-2.28) {Legend:};
\node[legend, fill=gray!12, anchor=west] (l1) at ([xshift=1.4mm]leglbl.east) {base table};
\node[legend, fill=white, anchor=west] (l2) at ([xshift=1.6mm]l1.east) {original model};
\node[legnew, anchor=west] (l3) at ([xshift=1.6mm]l2.east) {new model};
\node[legrew, anchor=west] (l4) at ([xshift=1.6mm]l3.east) {rewritten model};
\node[legend, fill=red!8, dashed, anchor=west] (l5) at ([xshift=1.6mm]l4.east) {pruned model};
\end{tikzpicture}%
}

%% file: sections/3_approach.tex
\section{Approach}
\label{sec:approach}

% Required preamble (in main.tex):
%   \usepackage{tikz}
% This file loads its own TikZ libraries below, so no \usetikzlibrary
% is needed in main.tex on its account.

\usetikzlibrary{arrows.meta, positioning, shapes.geometric, fit, calc}

\begin{figure*}[t]
\inv
\centering
\resizebox{\textwidth}{!}{%
\begin{tikzpicture}[
    font=\small,
    node distance=4mm and 6mm,
    stage/.style={
        draw, rounded corners=2pt, thick,
        minimum height=10mm, minimum width=30mm,
        align=center, fill=black!4,
        inner sep=2pt
    },
    freqstage/.style={
        draw, rounded corners=2pt, thick, dashed,
        minimum height=4mm, minimum width=28mm,
        align=center, fill=blue!5,
        inner sep=2pt
    },
    io/.style={
        align=center, font=\footnotesize\itshape, text=black!70
    },
    arrow/.style={
        -{Stealth[length=2.5mm]}, thick
    },
    freqarrow/.style={
        -{Stealth[length=2.5mm]}, thick, dashed, draw=blue!60!black
    },
    arrowlabel/.style={
        font=\scriptsize, midway, fill=white, inner sep=1pt
    }
]

% --- input ---
\node[io] (input) {SQL pipeline\\DAG $P$};

% --- four pipeline stages, left to right ---
\node[stage, right=of input] (s1) {Candidate\\Identification\\\scriptsize(Section~\ref{sec:candidate-identification})};
\node[stage, right=of s1]    (s2) {LLM\\Refactoring\\\scriptsize(Section~\ref{sec:llm-refactoring})};
\node[stage, right=of s2]    (s3) {Materialization\\Tuning\\\scriptsize(Section~\ref{sec:materialization})};
\node[stage, right=of s3]    (s4) {Global\\Selection\\\scriptsize(Section~\ref{sec:materialization})};

% --- output ---
\node[io, right=of s4] (output) {optimized\\DAG $P^\star$};

% --- main pipeline arrows with artifact labels ---
\draw[arrow] (input) -- (s1);
\draw[arrow] (s1) -- node[arrowlabel, above]{group} node[arrowlabel, below]{subgraphs} (s2);
\draw[arrow] (s2) -- node[arrowlabel, above]{verified} node[arrowlabel, below]{candidates} (s3);
\draw[arrow] (s3) -- node[arrowlabel, above]{per-cand.} node[arrowlabel, below]{deltas $\Delta_c$} (s4);
\draw[arrow] (s4) -- (output);

% --- frequency profile band: one wide box spanning under stages s1-s4 ---
\node[freqstage, below=2mm of s2.south,
      minimum width=125mm, minimum height=10mm,
      anchor=north] (fband)
  at ($(s1.south west)!0.5!(s4.south east) + (0,-5mm)$)
  {Frequency profile\\\scriptsize(source update rates + output demand,
   propagated through DAG; Section~\ref{sec:frequency})};

% Frequency arrows: profile feeds the three stages it touches, each labeled by layer
\draw[freqarrow] (fband.north -| s1.south) -- (s1.south)
  node[arrowlabel, right=1pt, fill=white]{\scriptsize Layer 3: hints};
\draw[freqarrow] (fband.north -| s3.south) -- (s3.south)
  node[arrowlabel, right=1pt, fill=white]{\scriptsize Layer 1: weights};
\draw[freqarrow] (fband.north -| s4.south) -- (s4.south)
  node[arrowlabel, right=1pt, fill=white]{\scriptsize Layer 2: reschedule};

% Final tuning loop-back: flat elbow (saves the vertical space the arc used)
\coordinate (rtlvl) at ($(s4.north)+(0,3.2mm)$);
\draw[arrow, dashed] (s4.north) -- (s4.north|-rtlvl)
  -- node[arrowlabel, above, fill=white]{final retune} (s3.north|-rtlvl)
  -- (s3.north);;

\end{tikzpicture}%
}
% \inv
\caption{End-to-end pipeline of \tool. The top row shows the four main stages and the artifacts that flow between them; after global selection, a final materialization pass retunes the combined DAG. The dashed band underneath is the frequency dimension (Section~\ref{sec:frequency}): a single frequency profile feeds three of the four stages, contributing structural splitting hints, cost weights for materialization tuning, and schedule corrections at selection. }
\label{fig:pipeline}
\inv
\end{figure*}
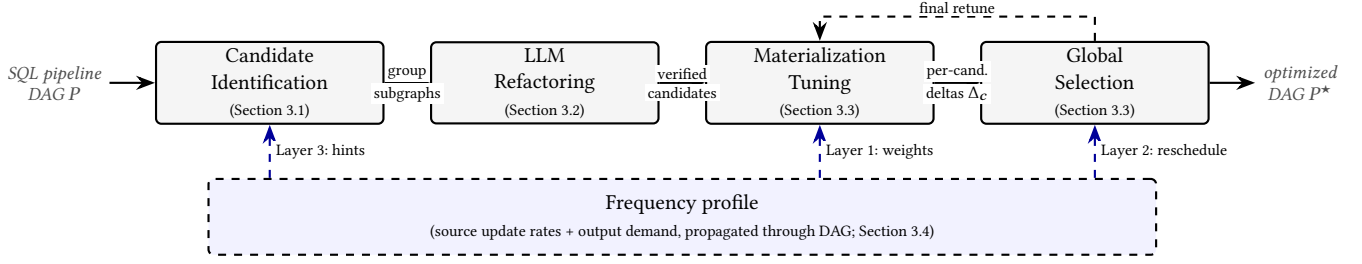

\tool{} implements the six optimization opportunities introduced in Section~\ref{sec:dependencies-as-signals} through three interacting implementation dimensions: graph refactoring, materialization tuning, and refresh-frequency tuning. Graph refactoring realizes dependency-edge simplification, non-local semantic reuse, downstream-aware pruning, and pipeline-aware work placement; materialization tuning realizes rewrite-materialization co-optimization by scoring each rewrite under its best table-or-view choices; and refresh-frequency tuning realizes frequency-aware optimization by aligning execution with both input-change rates and output-consumption rates. Global selection is a separate orchestration step that chooses a compatible subset of the candidates produced by these dimensions. Because our implementation and evaluation use dbt projects, this section uses \emph{model} to denote a SQL transformation node.
% \linc{If I remember correctly, Table 1 is just used for us to reference internally to make the terminology consistent, and we'll delete that eventually? If you're unsure, you can probably coordinate with Barzan.}
These dimensions are coupled: the best choice along one depends on the others, so they should not be optimized independently. We quantify these interactions with an ablation study in Section~\ref{sec:eval-attribution}.

% The benefit of a refactoring depends on which nodes are materialized, the best materialization configuration depends on the structure of the DAG, and both depend on how often each node executes. A pipeline with hundreds or thousands of models admits an enormous space of candidate refactorings, materialization assignments, and refresh schedules; exhaustive search is infeasible, and optimizing the three dimensions in separate passes misses these interactions and produces decisions that conflict at the boundaries.

%\linc{I think we need to reference Figure 5 at the beginning of this paragraph}
\tool{} realizes these dimensions in the pipeline of Figure~\ref{fig:pipeline}. Cheap structural analyses first scan the whole DAG and rank model groups by a fingerprint-based score that estimates the benefit of refactoring them jointly (Section~\ref{sec:candidate-identification}). The top-ranked groups are passed to an LLM, which reasons over each region and proposes one or more concrete refactorings (Section~\ref{sec:llm-refactoring}). 
The structural analyses decide where in the DAG to refactor; the LLM decides what refactoring to perform.
%\linc{Since we mentioned LLM, I think we need to add one sentence to talk about the correctness guarantee. Otherwise, I think that would be the first question that the reviewer raises.}
Every refactoring is adopted only after it passes data-backed equivalence checks run against the project's real data: a rewrite is rejected whenever its output diverges from the original's, so acceptance rests on observed agreement rather than the LLM's own assurance.
A refactoring's benefit depends on how its nodes are materialized, so \tool{} tunes each candidate's materialization with a learned cost model and selects a compatible subset to adopt (Section~\ref{sec:materialization}). The frequency-aware optimization affects all these stages (Section~\ref{sec:frequency}): it reweights the cost-driven decisions, removes refreshes that recompute unchanged data, and contributes structural splitting candidates using a frequency profile that combines source update rates with downstream output demand.

% % Required preamble (add to main.tex if not already there):
% % \usepackage{algorithm}
% % \usepackage{algpseudocode}

% \begin{algorithm}[t]
% \caption{\tool{} optimization pipeline.}
% \label{alg:pipeline}
% \begin{algorithmic}[1]
% \Require SQL pipeline DAG $P$, frequency profile $\Phi$
% \Ensure optimized DAG $P^\star$
% \State $f \gets \textsc{PropagateFrequencies}(P, \Phi)$
% \State $\mathcal{G} \gets \textsc{ReuseAnalysis}(P) \cup \textsc{PruneAnalysis}(P)$
% \Statex \hspace{3.6em} $\cup\ \textsc{FrequencySplitHints}(P, f)$ \Comment{\S\ref{sec:candidate-identification}}
% \State $\mathcal{C} \gets \textsc{LLMRefactor}(\mathcal{G}, P)$ \Comment{\S\ref{sec:llm-refactoring}}
% \State $\delta_{\text{orig}} \gets \textsc{MaterializationTune}(P, f)$ \Comment{\S\ref{sec:materialization}}
% \ForAll{$c \in \mathcal{C}$}
%   \State $P_c \gets \textsc{Apply}(P, c)$
%   \State $\Delta_c \gets \delta_{\text{orig}} - \textsc{MaterializationTune}(P_c, f)$
% \EndFor
% \State $H \gets \textsc{ConflictGraph}(\mathcal{C})$
% \State $\mathcal{C}^\star \gets \textsc{SelectILP}(\mathcal{C}, \{\Delta_c\}, H)$ \Comment{\S\ref{sec:materialization}}
% \State $P^\star \gets \textsc{Apply}(P, \mathcal{C}^\star)$
% \State $\textsc{MaterializationTune}(P^\star, f)$ \Comment{final retune}
% \State \Return $\textsc{ReconcileSchedule}(P^\star, f)$ \Comment{\S\ref{sec:frequency}}
% \end{algorithmic}
% \end{algorithm}

\subsection{Candidate Identification}
\label{sec:candidate-identification}

%\linc{I don't think we have established why we want to use LLM for refactoring yet. So I think either you have to establish that here, or just talk about refactoring is costly in general, without specifically emphasizing LLMs.}
Candidate identification finds where in the DAG  the four graph-refactoring optimizations from Section~\ref{sec:dependencies-as-signals} are worth pursuing.
\tool{} confines the refactoring to small, bounded regions rather than the whole project to ensure  the reasoning complexity over these interdependent models is tractable.
Since many regions may not be worth refactoring, \tool{} first scans the  DAG with cheap structural analyses that rank regions by likely benefit, reserving the expensive refactoring step (Section~\ref{sec:llm-refactoring}) for the few that pay off.

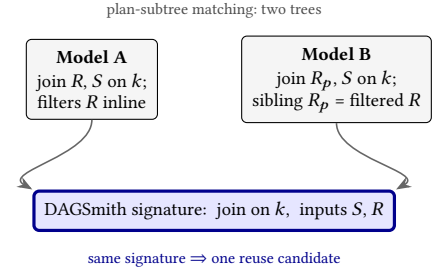
\begin{figure}[tb]
% \inv
\centering
\begin{tikzpicture}[
  font=\small,
  >={Stealth[length=2mm]},
  q/.style={draw, rounded corners=2pt, align=center, fill=black!4, inner sep=4pt, font=\footnotesize},
  canon/.style={draw=blue!55!black, very thick, rounded corners=2pt, align=center, fill=blue!10, inner sep=4pt, font=\footnotesize},
  arr/.style={->, semithick, draw=black!60},
  lbl/.style={font=\scriptsize, text=black!70}
]
\node[q] (A) {\textbf{Model A}\\[1pt]join $R$, $S$ on $k$;\\ filters $R$ inline};
\node[q, right=11mm of A] (B) {\textbf{Model B}\\[1pt]join $R_p$, $S$ on $k$;\\ sibling $R_p$ = filtered $R$};
% plan-matcher verdict above the boxes
\node[lbl, anchor=south] (top) at ($(A.north)!0.5!(B.north)+(0,1.5mm)$)
  {plan-subtree matching: two trees};
% signature below, leaving room for the curved arrows
\node[canon, anchor=north] (C) at ($(A.south)!0.5!(B.south)+(0,-9mm)$)
  {\tool{} signature:\; join on $k$,\; inputs $S$, $R$};
\draw[arr] (A.south) to[out=-90, in=150] (C.north west);
\draw[arr] (B.south) to[out=-90, in=30]  (C.north east);
\node[lbl, text=blue!55!black, below=1.5mm of C] {same signature $\Rightarrow$ one reuse candidate};
\end{tikzpicture}
% \inv
\caption{
Candidate identification discovers shared computation across models based on dataflow, not plan shape.
Models A and B compute the same join on $k$, but Model A filters $R$ inline while Model B reads the already-filtered $R$ from a sibling model $R_p$; on the compiled SQL the two are different trees over different tables, while \tool{} resolves the reference and reduces both to one signature and a single reuse candidate.}
\label{fig:dataflow-grouping}
\inv
\end{figure}

%\linc{Before getting into all the details below, I think it would be good to have one paragraph to discuss intuitively how your analysis works and what it achieves (if you can pair that with a figure, that's even better). The rest of this subsection is just implementation details. If intuitively the method makes sense, we would be good. We can either keep the details below or just condense them if we're short on space.}
The analysis must decide when two models perform the same expensive work, and the natural ways of testing for this fall short.
For example, matching on SQL text fails when the same logic is written differently. Matching on plan structure goes further—multi-query optimization and view-matching are already invariant to join order, predicate order, and aliasing—but it reasons about one compiled query at a time and treats a referenced model as an opaque input.
Therefore, it cannot recognize work shared across model boundaries. Figure~\ref{fig:dataflow-grouping} shows the problem: Models A and B compute the same join, but Model A filters its input inline while Model B reads the already-filtered input from a sibling model it references. On the compiled SQL, a plan-subtree matcher therefore sees two unrelated trees over different tables.

To address such challenges, DAGSmith matches on the operation each model computes via dataflow analysis, resolving references so that a model boundary no longer hides shared work, and reduces the two to a single reuse candidate, which the LLM and a data-backed equivalence check confirm is safe to share. It further performs two analyses: reuse finds repeated expensive work to share (non-local semantic reuse), and prune finds work that no downstream model consumes or that runs too early (dependency-edge simplification, downstream-aware pruning, and pipeline-aware work placement). Only the highest-scoring regions are passed to the LLM.

%At its core the analysis must decide when two models do the same expensive work, and the obvious tests miss real cases. Matching SQL text fails on any rephrasing. Matching plan structure is stronger, since multi-query optimization and view-matching already handle join order, predicate order, and aliasing, but it is keyed on the shape of one compiled query and treats a referenced model as an opaque input, so it misses work shared across model boundaries. In Figure~\ref{fig:dataflow-grouping}, Models A and B compute the same join, but Model A filters its input inline while Model B reads the already-filtered input from a sibling model it references, so on the compiled SQL a plan-subtree matcher sees two different trees over different tables.
%\tool{} instead matches on the operation each model computes, resolving references so that a model boundary does not hide shared work, and reduces the two to a single reuse candidate that the LLM and an equivalence check later confirm is safe to share. Two analyses work this way: \emph{reuse} finds repeated expensive work to share (opportunity~(2)), and \emph{prune} finds work that no downstream model consumes or that runs too early (opportunities~(1), (3), and~(4)). Only the highest-scoring regions go to the LLM.

\head{Dataflow representation}
Both analyses run on a \emph{dataflow representation} rather than on SQL text: \tool{} compiles each model into a tree of typed operators and resolves every column and \code{ref()} back through the models that produced it. Figure~\ref{fig:dataflow-grouping} shows why this is needed. Model B reads the sibling $R_p$, so at the SQL level its input is an opaque table name; resolving the reference substitutes $R_p$'s definition, the filtered $R$, and reconstructs the join B actually computes over $R$ and $S$. Model A reaches the same operation by filtering $R$ inline.
The two resolved trees coincide even though the \code{ref()} boundary hides the correspondence in the source, so the analyses can compare models by what they compute rather than by how the SQL is split across the DAG.

\head{Reuse analysis}
Reuse analysis summarizes each model's dataflow representation by a \emph{signature}: its expensive operations (joins and group-by aggregations), the keys they operate on, and their resolved inputs. Models whose signatures match are grouped into one \emph{reuse candidate}; the candidates are scored and ranked by the heuristic described below.
% Reuse analysis summarizes each model's dataflow representation by a \emph{signature}: its expensive operations (joins and group-by aggregations), the keys they operate on, and their resolved inputs. Models whose signatures match are grouped into one candidate and \textit{scored} by the cost of the shared operation and the number of models that share it. \linc{How the score is defined is still slightly vague. Can you be a bit more specific? Edit: see the comment at the end of this subsection first.} A high score is a \emph{targeting signal} that marks where the LLM should look.

\head{Prune analysis}
Prune analysis targets joins in two forms. A \emph{redundant} join produces a result no downstream model consumes, confirmed by tracing column lineage from the leaf models, and can be removed outright or, when it is a semi-join whose filter is already derivable from the probed relation, collapsed to a predicate. A \emph{mistimed} join is correct but computed too early, inflating an expensive downstream operation whose result does not depend on it; pushing it past that operation avoids carrying its product through the work. Both targets are different from local dead-code elimination in scope: whether a join's result is used, or used too soon, is a fact about the rest of the pipeline that a single-query rewriter cannot see. Each such join becomes a \emph{prune candidate}, scored and ranked by the same heuristic as reuse candidates.
% \linc{How does this analysis affect the ``score''? Is it independent of the score? Edit: see the comment in the next paragraph first.}

\head{From candidates to subgraphs}
Both analyses emit candidate model groups, ranked by a single score that estimates how much work a candidate would save. The score is the product of two factors: the cost of the candidate's expensive operation, approximated by the number of joins and group-bys it involves, and the number of models that benefit from it, namely the models that would share the computation for a reuse candidate, or the downstream models relieved of it for a prune candidate. \tool{} processes the groups from the highest score down, so the expensive LLM stage is reserved for the regions most likely to pay off. For each group, it extracts a subgraph of the candidate models and their immediate (one-hop) upstream parents, leaving out downstream consumers and siblings; the one-hop parents supply the inputs the next stage reasons over, while each candidate's output contract to consumers outside the subgraph stays fixed (Section~\ref{sec:llm-refactoring}). Subgraphs that exceed a node-count cap are skipped. The next subsection describes how the LLM turns these subgraphs into concrete refactorings.

% Both analyses produce a ranked stream of candidate model groups, scored by the same simple heuristic. The score multiplies a weight that counts the group's expensive operators (joins and group-bys) by the number of models that benefit: those that share the computation for reuse, and the downstream models relieved of it for pruning. \linc{okay, it seems we're re-introducing the definition of ``score'' again here, which is a bit confusing. I think either you don't mention the score earlier and only discuss it here, or, in the earlier analysis, you say the results of the analysis will be used to compute a ``score'' and forward-reference the scoring discussion here. TBH, I think the easiest way to illustrate this is still to have a figure that shows the relationship between these analyses and how they contribute to this ``score'', if you can do it...} \tool{} processes the groups from the highest score down. For each, it extracts a subgraph of the candidate models and their immediate (one-hop) upstream parents, leaving out downstream consumers and siblings; the one-hop parents supply the inputs the next stage reasons over, while each candidate's output contract to consumers outside the subgraph stays fixed (Section~\ref{sec:llm-refactoring}). Subgraphs that exceed a node-count cap are skipped. The next subsection describes how the LLM turns these subgraphs into concrete refactorings.

\subsection{LLM Refactoring}
\label{sec:llm-refactoring}

%\linc{Well, why do we want to use an LLM to do this in the first place? We either have to establish this earlier or justify it here.}
This stage produces the refactorings that realize the four graph-refactoring optimizations from Section~\ref{sec:dependencies-as-signals}. As motivated in the overview, producing these cross-boundary, semantic rewrites is a challenging reasoning task, which is why \tool{} uses state-of-the-art frontier LLMs rather than a fixed rule set.
% \linc{Say what kind of model this is (i.e., SOTA frontier models from commercial providers, etc.)} 
But asking an LLM to rewrite SQL that will run against a production warehouse raises three problems: the rewrite may be \emph{slower} than the original, may be \emph{incorrect}, or may be \emph{too expensive to generate} when a region contains many models.

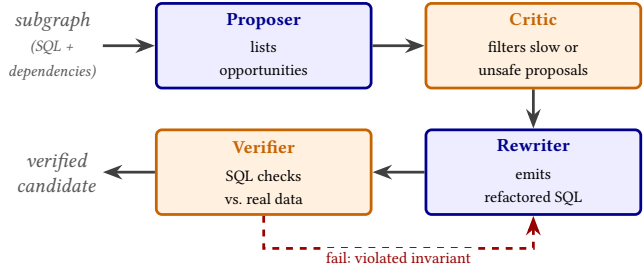
\begin{figure}[tb]
\inv
\centering
\resizebox{\columnwidth}{!}{%
\begin{tikzpicture}[
    font=\small,
    node distance=6mm and 8mm,
    rolebase/.style={draw, rounded corners=2.5pt, very thick,
        minimum height=13mm, minimum width=33mm, align=center, inner sep=3pt, font=\small},
    gen/.style={rolebase, fill=blue!7,   draw=blue!55!black},
    chk/.style={rolebase, fill=orange!12, draw=orange!78!black},
    io/.style={align=center, font=\itshape, text=black!65},
    arrow/.style={-{Stealth[length=3mm]}, very thick, draw=black!75},
    backarrow/.style={-{Stealth[length=3mm]}, very thick, dashed, draw=red!60!black},
]
\node[io] (in) {subgraph\\\footnotesize(SQL +\\\footnotesize dependencies)};
\node[gen, right=of in]   (prop) {{\color{blue!55!black}\textbf{Proposer}}\\[2pt]\footnotesize lists\\\footnotesize opportunities};
\node[chk, right=of prop] (crit) {{\color{orange!78!black}\textbf{Critic}}\\[2pt]\footnotesize filters slow or\\\footnotesize unsafe proposals};
\node[gen, below=of crit] (rew)  {{\color{blue!55!black}\textbf{Rewriter}}\\[2pt]\footnotesize emits\\\footnotesize refactored SQL};
\node[chk, left=of rew]   (ver)  {{\color{orange!78!black}\textbf{Verifier}}\\[2pt]\footnotesize SQL checks\\\footnotesize vs.\ real data};
\node[io, left=of ver] (out) {verified\\candidate};
\draw[arrow] (in)   -- (prop);
\draw[arrow] (prop) -- (crit);
\draw[arrow] (crit) -- (rew);
\draw[arrow] (rew)  -- (ver);
\draw[arrow] (ver)  -- (out);
\draw[backarrow] (ver.south) -- ([yshift=-5mm]ver.south) -| (rew.south);
\node[font=\footnotesize, anchor=north, fill=white, inner sep=1.5pt]
  at ($(ver.south)!0.5!(rew.south) + (0,-5mm)$) {\color{red!60!black}fail: violated invariant};
\end{tikzpicture}%
}
% \inv
\caption{\tool{}'s agentic refactoring loop. Four agents each fill one role: a \emph{proposer} and \emph{rewriter} (blue) generate, while a \emph{critic} and \emph{verifier} (amber) scrutinize. A failed equivalence check returns the violated invariant to the rewriter, which retries or reverts. }
\label{fig:llm-refactoring}
\inv
\end{figure}

% \linc{Fonts are too small. Need to reference the figure in Section 3.2 when we overview the solution. Probably also want to change the color/shape a bit or add some symbols. Right now, the figure looks a bit plain.}

% \linc{Maybe we can say we use an agentic framework to address this with four separate agents? That might be a bit better framing I feel like.}
\tool{} addresses all three with an agentic framework of four agents, each filling a single role  (Figure~\ref{fig:llm-refactoring}): a \emph{proposer} reads the region and lists candidate optimizations, an adversarial \emph{critic} rejects proposals that would slow the pipeline or silently change its results, a \emph{rewriter} turns the survivors into concrete SQL, and a \emph{verifier} tests each rewrite against the warehouse's real data before it is trusted. Separating proposal from criticism keeps a single prompt from having to be both creative and skeptical, and grounding acceptance in real data rather than in the model's own assurance is what makes adopting LLM-written SQL safe.

\head{Proposal}
The first call ingests the subgraph's SQL and dependency edges and produces a domain-level reading of the region: a short semantic summary of each model and the end-to-end business question the subgraph answers. This reading is what lets later stages recognize equivalent computations written in dissimilar SQL. 
The call then outputs a list of \emph{opportunities}, each describing the work it would remove or shrink and how—proposals to be rewritten and checked by later agents.

\head{Filtering} A second call acts as an adversarial critic, going through the proposals and trying to break them. We keep this separate from proposal on purpose: a single prompt asked to both propose and doubt does neither well, and in practice the proposing voice often wins out. The critic looks for two failure modes. The first is a slowdown—a proposal that looks like reuse-driven savings but adds a write and a scan without cutting downstream work. The second is a correctness trap, of which Figure~\ref{fig:critic-catch} shows one example: the proposal drops a join by reading a tag from a sibling deduplicated to one row per item, silently discarding the secondary tags the original join admitted. The critic returns one of three verdicts—\emph{approve}, \emph{reject}, or \emph{conditional}—where a conditional verdict records a data property the rewrite must satisfy, carried into the next stage.

\begin{figure}[!htbp]
\begin{minipage}[t]{0.48\linewidth}
\figsubheader{Original: join admits all tags}
\begin{lstlisting}[language=SQL]
SELECT i.item_id,
  MAX(CASE WHEN t.tag='A'
    THEN 1 ELSE 0 END) AS flag_a
FROM {{ ref('items') }} i
JOIN {{ ref('item_tags') }} t
  USING (item_id)
GROUP BY i.item_id
\end{lstlisting}
\end{minipage}
\hfill
\begin{minipage}[t]{0.48\linewidth}
\figsubheader{Proposal: drop join, read tag}
\begin{lstlisting}[language=SQL]
SELECT s.item_id,
  CASE WHEN s.tag='A'
    THEN 1 ELSE 0 END AS flag_a
FROM {{ ref('items_staged') }} s
-- items_staged keeps one tag/item;
-- secondary tags are silently lost
\end{lstlisting}
\end{minipage}
\inv
\caption{Correctness trap rejected by the critic. The original aggregates over all tags per item; the proposal reads from a sibling that has been deduplicated to one tag per item and silently drops the rest.}
\label{fig:critic-catch}
\inv
\end{figure}

\head{Rewriting}
The opportunities that survive filtering, together with the analysis and the SQL of the affected nodes, go to a third call that produces concrete refactored SQL: it may rewrite nodes, mark nodes as removed, or introduce new shared nodes. Conditional opportunities carry their recorded constraints into this prompt. A node with consumers outside the subgraph may be rewritten internally but must produce identical output rows, since its contract with those consumers is fixed.

%A rewrite that passed filtering can still change the data, especially when one bundles several opportunities and a trap rejected on its own reappears nested inside a larger change: the staged-input shortcut of Figure~\ref{fig:critic-catch}, for instance, can resurface inside a consolidated aggregation the critic approved on other grounds. \tool{} therefore treats equivalence as an empirical question against real data. A fourth call, the \emph{correctness critic}, states for each adopted opportunity the data invariant that must hold for the rewrite to be safe and renders it as a SQL diagnostic that returns zero rows when the invariant holds and rows when it is violated. The diagnostics run against the target warehouse under a per-query byte cap; any returned row is a counterexample drawn from real data. Failed invariants are appended to the rewriting prompt, and the LLM must either find a distinctly different strategy that respects them or restore the original logic. The loop ends when the rewrite passes all diagnostics, the LLM reports no further distinct strategy, or a per-group attempt budget is exhausted, in which case the group reverts to its original SQL and contributes no candidate.
\head{Equivalence verification}
Filtering judges proposals one at a time, but the rewriter often bundles several into a single change—and a trap that the critic would have rejected on its own can slip back in once it is buried inside a larger rewrite. The grain-collapse shortcut of Figure~\ref{fig:critic-catch}, for instance, can resurface inside a consolidated aggregation the critic approved on other grounds. 
Because a bundled rewrite can be wrong even after filtering, \tool{} checks each one against real data. A fourth call, the \emph{verifier}, writes a check for every adopted opportunity: a SQL query that returns rows only when the rewrite has changed the data. These run against the target warehouse under a per-query byte cap, so any row that comes back is a real counterexample. When a check fails, the violated invariant goes back to the rewriter, which must either find a different strategy that respects it or revert to the original. The loop stops once every check passes, the LLM has no new strategy to try, or the per-group attempt budget runs out—in which case the group keeps its original SQL and contributes nothing. Because these checks run on the project's own data, they reject any divergence that data exposes but do not amount to a formal equivalence proof, so \tool{} does not claim guaranteed correctness. Adopted rewrites are still meant for a final human check before deployment.

% \linc{Does this guarantee correctness? If not, do we have a manual verification stage? I think we could have a sentence to clarify this.}

\head{Model families and propagation}
A subgraph often contains a \emph{family} of near-duplicate models that apply the same operation to different data subsets, and running the full four-call sequence on each would be wasteful. \tool{} instead runs it once on a representative and propagates the accepted rewrite to the rest with a lighter per-member call. Where a project has no family structure, the sequence runs over every model unchanged.

\subsection{Materialization Tuning}
\label{sec:materialization}

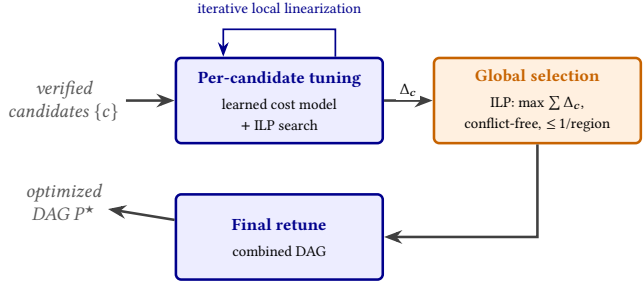
\begin{figure}[t]
\centering
\resizebox{\columnwidth}{!}{%
\begin{tikzpicture}[
    font=\small, node distance=8mm and 8mm,
    stagebase/.style={draw, rounded corners=2.5pt, very thick,
        minimum height=14mm, minimum width=34mm, align=center, inner sep=3pt, font=\small},
    tuneS/.style={stagebase, fill=blue!7,   draw=blue!55!black},
    selS/.style={stagebase, fill=orange!12, draw=orange!78!black},
    io/.style={align=center, font=\itshape, text=black!65},
    arrow/.style={-{Stealth[length=3mm]}, very thick, draw=black!75},
    looparrow/.style={-{Stealth[length=2.4mm]}, thick, draw=blue!55!black},
    albl/.style={font=\footnotesize, fill=white, inner sep=1.5pt},
]
\node[io] (in) {verified\\candidates $\{c\}$};
\node[tuneS, right=of in] (tune) {{\color{blue!55!black}\textbf{Per-candidate tuning}}\\[2pt]\footnotesize learned cost model\\\footnotesize $+$ ILP search};
\node[selS, right=of tune] (sel) {{\color{orange!78!black}\textbf{Global selection}}\\[2pt]\footnotesize ILP: max $\sum\Delta_c$,\\\footnotesize conflict-free, $\le\!1$/region};
\node[tuneS, below=of tune] (re) {{\color{blue!55!black}\textbf{Final retune}}\\[2pt]\footnotesize combined DAG};
\node[io, below=of in] (out) {optimized\\DAG $P^\star$};
\draw[arrow] (in)   -- (tune);
\draw[arrow] (tune) -- node[albl,above]{$\Delta_c$} (sel);
\draw[arrow] (sel.south) |- (re.east);
\draw[arrow] (re)   -- (out);
% elbow self-loop over the tuning box: iterative local linearization
\draw[looparrow] ([xshift=-8mm]tune.north east) -- ++(0,4.5mm) -| ([xshift=8mm]tune.north west);
\node[font=\footnotesize, anchor=south, inner sep=1pt, text=blue!55!black]
  at ([yshift=6.8mm]tune.north) {iterative local linearization};
\end{tikzpicture}%
}
% \inv
\caption{\tool{}'s materialization tuning and global selection.}
\label{fig:materialization}
\inv
\end{figure}

% \linc{Fonts are too small. Need to reference the figure in Section 3.3 when we overview the solution. Probably also want to change the color/shape a bit or add some symbols. Right now, the figure looks a bit plain.}

This stage realizes rewrite-materialization co-optimization. 
A refactoring's benefit cannot be determined in isolation: whether it helps depends on which of its nodes are stored as tables and which are inlined as views, and the best choice differs from one candidate to the next, so a candidate must be scored together with its own best materialization. This is especially important when a rewrite creates additional reusable output: the richer rewrite may be locally more expensive, but it can become beneficial when the output is materialized once and reused by enough downstream consumers. Even then, a second challenge remains: candidates overlap and cannot all be adopted at once, so they must be composed into a compatible set. This subsection addresses both (Figure~\ref{fig:materialization}): tuning one candidate's materialization, then selecting a compatible set.

\head{Tuning one candidate}
To score a candidate, \tool{} must find its best materialization: the assignment of each node to a table or a view that makes the whole pipeline cheapest to run, where \emph{cost} is 
the total slot time of executing the entire pipeline once.
% \linc{of the entire project? Or just one model?}
 That lowest cost becomes the candidate's score.

 Finding that assignment is hard because the table-or-view choice is not node-local: a view re-executes its SQL into every consumer that reads it, recursively through any view-parent. Flipping a node to a view therefore pushes its work downstream, coupling its cost to its descendants, so total cost is not a sum of per-node costs.
 
% Finding that assignment is hard because the table-or-view choice is not node-local. A table is a barrier: it is built once, and its consumers read its stored rows. A view is not: reading a view re-executes its SQL, expanding recursively through any view-parent until it reaches a table or source. Flipping a node from a table to a view therefore drops its own build cost but pushes its work, together with the view chain above it, into every consumer that reads it, once per consumer execution. A node's cost thus depends on how its ancestors are materialized, so the choices are coupled and total cost is not a sum of per-node costs. 

% \linc{This paragraph seems a bit abrupt. It doesn't connect either the paragraph above or below. What's the point of this paragraph? How is it relevant to the paragraph above and below?}
 
This coupling creates two challenges that \tool{} handles in turn. 
% \linc{What are the two difficulties? The ``two reasons'' mentioned two paragraphs ago? Or something from the last paragraph?} 
The first is evaluation: scoring an assignment typically involves executing the pipeline, so \tool{} instead uses a learned cost model to enable efficient approximation. Two robust (Huber~\cite{huber1964robust}) regressors, fit on instrumented runs of the original pipeline, predict each node's output cardinality and its slot time from structural features parsed off its compiled SQL, each conditioned on the predicted cardinalities of the node's parents.
Because a view's work folds into whoever reads it, \tool{} accumulates each node's features up its view chain. A node's predicted cost then reflects the work it actually performs under the current table-or-view assignment, not just its own SQL. New nodes a refactoring introduces have no measured runs of their own and are predicted the same way, from their SQL features and their parents' predicted cardinalities.
% Because a view's work folds into whoever reads it, these features are accumulated along each view chain, so a node is priced on the work it performs under the assignment in question; new nodes a refactoring introduces, having no measurements of their own, are predicted the same way from their SQL and their parents' cardinality. 
% \linc{$\leftarrow$ The earlier sentence is difficult for me to parse.} 
The model only has to rank assignments within one project rather than predict absolute runtimes, so directional accuracy is enough, and the Huber loss keeps a handful of very large queries from dominating the fit.
 
The second is search: the best assignment must be found among exponentially many ($2^n$ for a candidate of $n$ nodes) possible solutions. \tool{} navigates this with \textbf{iterative local linearization}~\cite{griffith1961slp,palaciosgomez1982slp}, a successive-approximation scheme that replaces the true nonlinear cost with a local linear surrogate, solves it, and re-linearizes at the result. Concretely, at iteration $k$, holding the current assignment $t^{(k)}$ fixed, \tool{} uses the cost model to measure two local quantities: a baseline build cost $B^{(k)}_j$ for each node $j$ (its direct parents forced to tables) and an edge-local delta $U^{(k)}_{ij}$, the extra cost at $j$ when only parent $i$ is inlined as a view. Treating these as constants for the iteration, it solves a small ILP for $t^{(k+1)}$:
\begin{equation*}
\min_{t,\,y}\; \sum_{j \in V} t_j\, B^{(k)}_j \;+\; \sum_{i \to j \in E} y_{ij}\, U^{(k)}_{ij}, \qquad \text{s.t.}\quad y_{ij} = (1 - t_i) \wedge t_j,
\end{equation*}
where $y_{ij}$ is $1$ exactly when node $j$ is built and reads a view-parent $i$, and a trust-region cap on flips keeps each step where the coefficients hold. 
Because flipping a node changes the cardinalities its descendants see, $B$ and $U$ are recomputed after each move and the step repeated until the assignment stops changing; an effect the current coefficients miss, such as the cost of also flipping a sibling, then surfaces in the next round. This is what makes the coupled, nonlinear assignment tractable: instead of enumerating the $2^n$ assignments, each round solves one small ILP that weighs all nodes' table-or-view choices jointly under the current cost estimates, capturing interactions that a greedy, one-node-at-a-time search would miss. The search reaches a local optimum, not a global one, but this is safe: the original DAG is tuned by the same procedure and kept as a fallback, so an under-tuned candidate simply loses to it, costing at most a missed improvement rather than a regression below the original.

\head{Selecting a compatible set}
Tuning leaves each candidate with a single number, its \emph{delta}: how much cheaper the pipeline runs if that one refactoring is applied, with the original and the refactored version each at its own best materialization. Selection then decides which candidates to adopt. It would take all of them if it could, but two things get in the way. First, refactorings can conflict: two that rewrite, create, or remove the same model cannot both be applied. Second, the LLM sometimes proposes several alternative refactorings for the same region, and at most one of them may be chosen. \tool{} therefore looks for the set of candidates with the largest total delta that has no conflicts and at most one variant per region, a small problem it hands to an off-the-shelf integer-program solver. One correction remains: because each delta was measured with the candidate tuned on its own, the best table-or-view choices for the chosen combination need not match any single candidate's, so \tool{} retunes the combined project one final time, rerunning the same cost-model-guided search used for each candidate, now over the merged DAG with all selected refactorings in place. 
% \linc{How this ``retune'' is done? Maybe briefly explain this with one/half a sentence.}

\subsection{Frequency-Aware Optimization}
\label{sec:frequency}

This stage realizes frequency-aware optimization. The stages so far treat each node's slot time as a fixed per-execution cost. In a recurring pipeline that is suboptimal: a node's real overhead is its per-run cost multiplied by how often it must actually run. Treating cost as per-execution therefore overvalues savings on rarely-run models and undervalues them on frequently-run ones. \tool{} adds a frequency dimension to correct this. Effective frequencies are derived from the declared schedule at sink nodes and the observed update rates at source tables, propagated through the DAG, and applied at three layers;  the profile captures both how often inputs change and how often downstream outputs are consumed.  The first two layers leave the DAG structure intact, while the third feeds a new class of candidates back to candidate identification (Section~\ref{sec:candidate-identification}).

\head{Layer 1: Frequency as cost weight}
Materialization tuning and selection (Section~\ref{sec:materialization}) minimize cost as if every model ran at the same rate. \tool{} reweights their cost-model inputs by each node's effective frequency before the decisions are made; nothing else about them changes.

\head{Layer 2: Reconciling scheduled and effective frequency}
A model's effective frequency is the rate at which it produces \emph{new} output, the slower of how often its consumers need fresh output ($f^{\text{demand}}_j$) and how often its inputs actually change ($f^{\text{data}}_j$):
\begin{equation*}
f_j = \min\!\big(f^{\text{demand}}_j,\; f^{\text{data}}_j\big).
\end{equation*}
Where the schedule fires faster than $f_j$, the model re-executes and reproduces the output it produced last time. \tool{} flags such over-scheduled models and reduces their schedule to $f_j$. This changes neither SQL nor materialization, only the schedule, yet every redundant execution it removes is eliminated outright.

\head{Layer 3: Frequency-driven splitting hints}
The first two layers leave the DAG structure intact; the third changes it. When a model has parents that change at different rates, it inherits the fastest, so any work inside it that depends only on slow-changing parents is recomputed at the fast rate even though its inputs have not changed. \tool{} identifies such models and emits the location to candidate identification as a hint, alongside the reuse and prune signals. The LLM then proposes a frequency split of the kind shown in Figure~\ref{fig:strategy-c-sql-frequency}: the slow-side work is extracted into its own slow-cadence model and the original becomes a lightweight join, after which the candidate flows through the same filtering, rewriting, verification, and selection as any other.

%% file: sections/4_eval.tex
\section{Evaluation}
\label{sec:eval}

We evaluate \tool{} on real dbt projects, organized around three questions:
\begin{enumerate}
    \item How does \tool{} compare to existing baselines for SQL pipeline optimization, and what classes of optimization account for the gap?
    \item How much does each component of \tool{}'s architecture contribute to the overall gain? We ablate the three  implementation dimensions  (refactoring, materialization, and frequency) and the dataflow-driven targeting that feeds the LLM stage.
    \item Is the underlying machinery sound and affordable? We measure  the effectiveness of the equivalence-verification loop at catching unsafe rewrites, the LLM cost of running the system end-to-end, and the accuracy of the cost model that drives materialization decisions.
\end{enumerate}

\head{Summary of results}
On the open-source Tuva and Stripe dbt projects, \tool{} reduces BigQuery slot-time, the unit billed under capacity pricing, by 51.1\% and 47.4\% to build the pipeline once, and by 67.7\% and 54.8\% in recurring cost averaged across three representative refresh schedules per project. Both figures beat the strongest baseline by a wide margin (on Tuva, 67.7\% against its 37.9\% recurring-cost reduction), because \tool{} reaches cross-model refactoring, materialization revision, and frequency-aware rescheduling that single-query and syntactic cross-model rewriters cannot. An ablation confirms the three  implementation dimensions are complementary, and that dataflow-guided targeting spends a fixed LLM budget more effectively than random subgraph selection (51.1\% vs.\ 43.1\% single-run). 
The supporting machinery is sound and affordable (Section 4.4): the verification loop reliably rejects unsafe rewrites, the LLM cost is paid once and amortized over many runs, and the learned cost model is accurate enough for the decisions it drives.

\subsection{Experimental Setup}
\label{sec:eval-setup}

\head{Workloads}
We evaluate on two real dbt projects. \emph{Tuva} is an open-source dbt package for healthcare data transformation. The full project comprises 842 SQL models organized across 17 modules. We use the claims preprocessing module, which contains 255 models (about 30\% of the project) and is the subset whose models exhibit substantial computational complexity. We run it on a synthetic patient population of 10K, with the largest source table at 505\,MB. \emph{Stripe} is a payments-analytics project of 65 models, already factored by hand into intermediate stages so that the easy syntactic wins have been taken. We run it with the largest source table at 199\,MB. The two projects together cover the range we want to evaluate: a large project that has accumulated cross-model patterns from multiple analysts, and a smaller, already-factored project where remaining gains must come from materialization, frequency, or semantic restructuring beyond syntactic deduplication.

\head{Testbed}
Experiments run on BigQuery with a 100-slot reservation on the Standard edition under capacity-based pricing.
\tool{} is not specific to BigQuery: it optimizes SQL pipelines in general, and we use BigQuery because our open-source workloads are built on them. We report slot-time as the primary cost metric, since it is the billing unit under capacity pricing, and elapsed time as a secondary latency metric. Each reported number is the median of three runs. Every LLM call in \tool{} uses OpenAI's GPT-5.4. To simulate realistic production schedules, we use three frequency profiles per project that assign root and leaf models to hourly, daily, or weekly cadences, with interior models derived by propagation. Within the root and leaf sets, cadences are assigned at random, with
leaves skewed faster than roots, mirroring production schedules, which
are driven by consumer demand at the sinks rather than the true update
rate at the sources. This mismatch is exactly what the frequency
dimension exploits. Table~\ref{tab:freq-profiles} summarizes the average cadence mix per project; all frequency-weighted results in this section are averaged across the three profiles.

% \begin{table}[tb]
% \centering
% \caption{Frequency profiles. Each row is the fraction of root and leaf models at each cadence.}
% \label{tab:freq-profiles}
% \small
% \begin{tabular}{llrrrrrr}
% \toprule
% & & \multicolumn{3}{c}{Roots} & \multicolumn{3}{c}{Leaves} \\
% \cmidrule(lr){3-5} \cmidrule(lr){6-8}
% Project & Profile & wk & day & hr & wk & day & hr \\
% \midrule
% \multirow{4}{*}{Tuva}
% & batch-heavy        & 0.07 & 0.71 & 0.21 & 0.05 & 0.15 & 0.80 \\
% & moderate-mismatch  & 0.18 & 0.57 & 0.25 & 0.05 & 0.20 & 0.75 \\
% & mixed-cadence      & 0.29 & 0.43 & 0.29 & 0.10 & 0.25 & 0.65 \\
% & \textbf{average}   & \textbf{0.18} & \textbf{0.57} & \textbf{0.25} & \textbf{0.07} & \textbf{0.20} & \textbf{0.73} \\
% \midrule
% \multirow{4}{*}{Stripe}
% & steady-sync         & 0.10 & 0.70 & 0.20 & 0.05 & 0.55 & 0.40 \\
% & high-frequency      & 0.05 & 0.50 & 0.45 & 0.00 & 0.45 & 0.55 \\
% & conservative-batch  & 0.25 & 0.60 & 0.15 & 0.10 & 0.55 & 0.35 \\
% & \textbf{average}    & \textbf{0.13} & \textbf{0.60} & \textbf{0.27} & \textbf{0.05} & \textbf{0.52} & \textbf{0.43} \\
% \bottomrule
% \end{tabular}
% \end{table}

% ===== Compact version: averages only. Uncomment to swap in. =====
\begin{table}[tb]
\centering
\caption{Frequency profile averages per project. Each row is the fraction of root and leaf models at each cadence, averaged across three profiles.}
% \inv
\label{tab:freq-profiles}
\small
\begin{tabular}{lrrrrrr}
\toprule
& \multicolumn{3}{c}{Roots} & \multicolumn{3}{c}{Leaves} \\
\cmidrule(lr){2-4} \cmidrule(lr){5-7}
Project & wk & day & hr & wk & day & hr \\
\midrule
Tuva   & 0.18 & 0.57 & 0.25 & 0.07 & 0.20 & 0.73 \\
Stripe & 0.13 & 0.60 & 0.27 & 0.05 & 0.52 & 0.43 \\
\bottomrule
\end{tabular}
% \inv
\end{table}

\head{Cost metrics}
We report two views of cost. \emph{Single-run cost} is the cost of executing the full DAG once. \emph{Frequency-weighted cost} weights each model's per-execution cost by the rate at which it actually needs to run, summed over the DAG; under capacity pricing this corresponds to the project's billed cost over a period. Refactoring and materialization affect both metrics; frequency optimization affects only the frequency-weighted metric. Improvements are reported as percent reduction relative to the original project.

\head{Baselines}
We compare against three baselines spanning the design space of SQL pipeline optimization.

\emph{LearnedRewrite}~\cite{learnedrewrite} is a rule-based per-query rewriter. It applies a portfolio of equivalence-preserving transformations to each compiled SQL independently, guided by a learned policy over which rules to apply and in what order. It is competitive on standalone analytical queries but has no view of the surrounding DAG.

\emph{GenRewrite}~\cite{genrewrite} is an LLM-based per-query rewriter. It introduces natural-language rewrite rules that transfer knowledge across queries and a counterexample-guided loop that iteratively corrects syntactic and semantic errors in the rewrites. Like LearnedRewrite, it operates one model at a time.

\emph{CSE} is a cross-model common subexpression baseline~\cite{scopeCSE}, which detects subexpressions shared across a SCOPE script's stages and materializes each once, re-optimized at the query optimizer's memo. The memo step is not portable, since managed warehouses do not expose their optimizer's internal state to external tools. We adapt the approach by detecting shared subexpressions at the AST level and selecting which to materialize heuristically. It is the only baseline
in our comparison with cross-model scope, and represents the best a purely
syntactic method can achieve on a managed warehouse without access to
optimizer internals.

\subsection{Comparison with Baselines}
\label{sec:eval-baselines}

Table~\ref{tab:overall} reports each method's improvement over the original project on Tuva and Stripe; \tool{} leads on every metric on both. The gap between the single-run and frequency-weighted columns is the contribution of frequency optimization, which affects only the frequency-weighted view.

\begin{table}[tb]
\centering
\caption{Improvement over the original project (\% reduction; higher is better). }
\inv
\label{tab:overall}
\small
\begin{tabular}{lrrrr}
\toprule
& \multicolumn{2}{c}{Single-run} & \multicolumn{2}{c}{Freq-weighted} \\
\cmidrule(lr){2-3} \cmidrule(lr){4-5}
& Elapsed & Slot & Elapsed & Slot \\
\midrule
\multicolumn{5}{l}{\emph{Tuva (255 models)}} \\
LearnedRewrite & $-19.6$ & $-2.5$ & $-20.2$ & $-3.0$ \\
GenRewrite     & $9.7$   & $17.7$ & $21.5$  & $15.1$ \\
CSE            & $12.2$  & $34.6$ & $27.7$  & $37.9$ \\
\tool{}        & $\mathbf{30.9}$ & $\mathbf{51.1}$ & $\mathbf{42.6}$ & $\mathbf{67.7}$ \\
\midrule
\multicolumn{5}{l}{\emph{Stripe (65 models)}} \\
LearnedRewrite & $2.6$   & $2.7$   & $-0.5$  & $4.5$  \\
GenRewrite     & $7.6$   & $7.1$   & $1.1$   & $9.3$  \\
CSE            & $-3.2$  & $-5.6$  & $-3.1$  & $-0.6$ \\
\tool{}        & $\mathbf{40.0}$ & $\mathbf{47.4}$ & $\mathbf{36.1}$ & $\mathbf{54.8}$ \\
\bottomrule
\end{tabular}
\inv
\end{table}

Each baseline falls short for a different structural reason. We name the limit in each case, then summarize the classes of optimization that account for the gap above the strongest baseline.

\head{LearnedRewrite: no net improvement on either project} LearnedRewrite applies equivalence-preserving rules independently to each model and therefore cannot reach any of the cross-model optimizations identified in Section~\ref{sec:dependencies-as-signals}, such as non-local semantic reuse and downstream-aware pruning, whose benefit depends on neighboring models. On Tuva, the one rule that fires at scale replaces a post-join \code{SELECT DISTINCT} with a \code{GROUP BY} that pre-deduplicates each join input. When the join key is near-unique, this grouping is nearly free and lets the query skip an expensive \code{DISTINCT} over the wide joined result, making the rewrite faster. When many rows share each key value, the warehouse already runs the original efficiently as a semi-join, and the extra grouping is wasted work that slows the rewrite down. Across the models the rule touches, wins and losses are roughly balanced. On Stripe, which is already hand-factored, LearnedRewrite's rules find little to match and it changes almost nothing. Overall, LearnedRewrite barely changes either project's cost: rule-based per-query rewriting has little left to do once the warehouse planner has already handled the simple per-query cases, and the changes that would actually help these projects span multiple models, which it cannot do.

\head{GenRewrite: cuts cost within a single model, not across the pipeline} GenRewrite restructures a single model to compute the same result more cheaply, such as replacing an $O(n^2)$ self-join with $O(n)$ window functions, a rewrite current rule-based systems do not support; on Tuva these account for its 17.7\% slot-time reduction. The gain is bounded because every rewrite stays inside a single model: GenRewrite cannot share work across models, change materialization, or skip refreshes. It can also regress, because it rewrites each model on its own. The same expensive self-join appears in several near-identical models. It applied the O(n) rewrite to one model but, on a near-identical sibling, produced three self-joins costing over twice the original, unable to recognize the two as the same. On Stripe, already hand-factored, fewer such rewrites are available and the gain is smaller (a 7.1\% slot-time reduction). 
% GenRewrite optimizes each model without seeing the others, which both caps its savings and lets it regress sometimes. 

\head{CSE: syntactic matching sees duplication, not equivalence} CSE matches subexpressions across models by their SQL structure, and is the only baseline that works across models. Where the same SQL recurs, it does well: Tuva has a large family of models that each scan and classify the same claims source with nearly identical SQL, and CSE detects the repeated work and runs it once for the whole family, which is most of its 34.6\% Tuva slot-time reduction. Where the SQL differs, it cannot recognize the reuse: six models compute the same per-encounter totals, each from different inputs and in different SQL, and CSE does not see them as the same, recovering little of the shared work. On Stripe, CSE slightly increases cost rather than reducing it, the only baseline to do so. Extracting a subexpression into a shared model only pays off when enough consumers reuse it; on Stripe, what CSE extracts is reused by too few consumers to cover the cost of building and scanning it.

\head{\tool{}: four patterns that the baselines structurally cannot reach} \tool{}'s advantage over the strongest baseline comes from four recurring patterns. The first two are cross-model refactorings that go beyond matching SQL text; the last two are materialization choices that no SQL rewriter can change. We give one example of each from the workloads.

\emph{(1) Sharing equivalent work that is written in different SQL.} Models that compute the same thing over different inputs are folded into one shared model, even when their SQL does not match. Take the six encounter models from above, which compute the same per-encounter totals but cannot be matched on SQL text: \tool{} builds one shared rollup that carries the encounter type as a column and replaces all six with short lookups, cutting their slot-time by about 77\%, against CSE's 14\%.  CSE detects that the six share structure but skips them, because their differing encounter-type literals leave the SQL not byte-identical; the equivalence is clear only once the LLM reads what each model computes.

\emph{(2) Replacing repeated wide-table joins with a narrow shared intermediate.} In one Tuva consumer, a union of 32 branches, seven of them each join a different encounter model to the same wide claim-line staging table only to read two of its columns. \tool{} materializes those two columns once as a narrow intermediate and points all seven joins at it, cutting the consumer's slot-time by about 74\%. The saving comes from the joins, not the scan: probing a two-column intermediate builds a far smaller hash table than the wide staging table (the warehouse already prunes unused columns). CSE cannot find this, because each branch joins a different upstream model and no two branches are the same SQL fragment to match.

\emph{(3) Storing a staging chain as views so the planner can optimize across it.} A staging model stored as a table is built in full and read back as a fixed result the planner cannot optimize across. Stored as a view, it is just SQL the planner folds into the query that reads it, letting it push that query's filters down to the staging's sources and drop intermediate work the query never uses. \tool{} treats this table-or-view choice as something to tune. On Stripe, switching the fifteen staging models behind a wide reporting model from tables to views, with no change to any SQL, cut the data it processes by about a third and its slot-time by about 60\%. Materializing each staging as a table seems like it should help by computing it once, but a table is built in full before the reporting query's filters can apply, leaving the planner nothing to trim; as views, those filters reach the sources, and the whole query is optimized at once. The table-or-view choice is a configuration setting, not part of the SQL; a tool that only rewrites compiled SQL never sees it.

\emph{(4) Removing intermediates that feed only one model.} On Tuva, six normalize-and-vote stages were each materialized as a table, even though each fed only a single downstream model. Materializing an intermediate pays off only when several models reuse it; for a single consumer, the build cost is wasted. \tool{}'s tuner flips these single-consumer stages to views, folding each into its one consumer and dropping the family's slot-time by about 24\%.  CSE keeps all six as tables and gets worse on the same family, because it cannot change materialization at all. No SQL rewriter can do this, because the choice is a materialization setting, not part of the SQL.

% Frequency optimization accounts for the gap between the single-run and frequency-weighted columns of Table~\ref{tab:overall}. It works in three ways: it weights each model's cost by how often it actually runs, focusing tuning on the models that run most; it cuts the schedule of models that run more often than their inputs change, removing redundant reruns; and it splits slow-changing work out of fast-refreshing models, moving that work onto a slower cadence. \linc{there are other layers in frequency-aware optimization, right? Did they help? I think we want to say a bit more here, since we have a full subsection on this optimization} \jie{to be done, will fine a case study later.} Like materialization, none of this is visible to a tool that sees only the compiled SQL.

Frequency optimization accounts for the gap between the single-run and frequency-weighted columns of Table~\ref{tab:overall}, realized through the three layers of Section~\ref{sec:frequency}. The first two carry most of it: reweighting each model's cost by how often it actually runs steers tuning toward the models that dominate the recurring bill, and lowering the schedule of models that fire faster than their inputs change removes redundant reruns outright. The third layer, splitting slow-changing work out of a fast-refreshing model, is the most structural change and adds a modest 3.6\% reduction, because splittable models are rarer than they appear. Splitting needs both parents that change at different rates and a slow side separable from the fast path, and most models fail one of the two: all their inputs change at the same rate, or the slow and fast inputs are entangled at the row level. It pays off only when the slow side merely appends columns, as in a left join that adds a lookup description. Like materialization, none of this is visible to a tool that sees only the compiled SQL. 
% \linc{This paragraph is getting slightly long. Can condense a bit if needing space.}

The two projects lean on these patterns differently: \tool{}'s advantage on Tuva comes mostly from patterns 1 and 2 (cross-model refactoring), and on Stripe from pattern 3 (materialization). The same pipeline produces both because it applies whatever fits each project rather than relying on a single mechanism. The next subsection quantifies each component's contribution.

\subsection{Component Ablation}
\label{sec:eval-attribution}

Section~\ref{sec:eval-baselines} described the \emph{classes of optimization} responsible for \tool{}'s gap above the baselines. This subsection asks how much each of \tool{}'s own architectural components contributes and how they interact.
We ablate along two axes: the three implementation dimensions of LLM refactoring (R; Section~\ref{sec:llm-refactoring}), materialization tuning (M; Section~\ref{sec:materialization}), and frequency optimization (F; Section~\ref{sec:frequency}); and the dataflow analysis that targets the LLM stage (Section~\ref{sec:candidate-identification}). We ablate on Tuva alone, since it is the larger and more structurally varied project and exercises all three dimensions meaningfully, whereas on the smaller Stripe project, attributing gains to individual components is unreliable. 

\subsubsection{Three Optimization Dimensions}
\label{sec:eval-ablation-rmf}

Table~\ref{tab:ablation-rmf} reports the improvement of each subset of $\{R, M, F\}$ over the original Tuva project.

\begin{table}[tb]
\centering
\caption{Tuva: dimension ablation (\% reduction over the original project, averaged across the three Tuva profiles in Table~\ref{tab:freq-profiles}). R = refactoring, M = materialization tuning, F = frequency optimization. Bold marks the best variant in each column. }
\inv
\label{tab:ablation-rmf}
\small
\begin{tabular}{lrr}
\toprule
Variant & Elapsed & Slot \\
\midrule
original         & --              & --              \\
F only           & $13.2$          & $11.0$          \\
M only           & $34.7$          & $37.0$          \\
R only           & $40.0$          & $49.1$          \\
M + F            & $31.0$          & $54.0$          \\
R + M            & $\mathbf{43.1}$          & $54.7$          \\
R + M + F (full) & $42.6$ & $\mathbf{67.7}$ \\
\bottomrule
\end{tabular}
\inv
\end{table}

% The numerics admit four observations:
% , each reflecting a property of how the dimensions are designed to interact.

\head{Refactoring and materialization mostly remove work parallel to the critical path}  Recall that slot measures the total CPU-time the project consumes, while elapsed measures wall-clock time from the first model to the last. For every variant that includes refactoring or materialization, the slot reduction exceeds the elapsed reduction (R-only, for instance, cuts slot by 49.1\% but elapsed by 40.0\%), because \tool{} removes work that runs parallel to the critical path more readily than work on it: consolidating many sibling models into one shared intermediate saves slot proportional to the number of siblings but shortens elapsed only by the slowest sibling's share. Frequency optimization is the exception, cutting elapsed and slot by comparable amounts (13.2\% vs.\ 11.0\%), because it removes whole scheduled executions rather than parallel work. The same mechanism explains the full pipeline: adding F on top of R+M raises the slot reduction sharply (54.7\% to 67.7\%) while elapsed is essentially unchanged (43.1\% to 42.6\%, a difference within run-to-run noise), because the redundant executions F eliminates lie off the critical path. F thus lowers billed cost without affecting wall-clock latency, which is exactly its intended effect.

% \head{Slot reductions track total CPU work, elapsed tracks the critical path} \linc{The paragraph head should summarize what we have learned about your approach (or different components), instead of which metrics track which.} Recall that slot measures the total CPU-time the project consumes; elapsed measures the wall-clock time from first model to last. For every variant that includes refactoring or materialization, the slot reduction exceeds the elapsed reduction, because \tool{} removes work parallel to the critical path more aggressively than work on it: consolidating many sibling models into one shared intermediate saves slot proportional to the sibling count but saves elapsed only the slowest sibling's share. The gap is largest for R-only (49.1\% slot vs.\ 40.0\% elapsed). Frequency optimization is the exception, cutting elapsed and slot by comparable amounts (13.2\% vs.\ 11.0\%) because it removes whole scheduled executions rather than parallel work. This is why the full pipeline holds the best slot number but not the best elapsed: on top of R+M, F adds heavily to slot (54.7\% to 67.7\%) by removing off-critical-path runs while leaving elapsed flat (43.1\% to 42.6\%), a half-point difference within the run-to-run noise band against a 13-point slot gain that is F's intended effect.

% Percentages are of the 255 original models.

\begin{table}[tb]
\centering
\caption{\tool{}'s coverage of Tuva's 255 models. }
\inv
\label{tab:coverage}
\small
\begin{tabular}{lr}
\toprule
Category & Count \\
\midrule
SQL refactored only              & 90 (35.3\%) \\
Materialization retuned only     & 13 \phantom{(0}(5.1\%) \\
Both refactored \& retuned       &  3 \phantom{(0}(1.2\%) \\
New models created               & 11 \phantom{(0}(4.3\%) \\
\midrule
Total models modified            & 117 (45.9\%) \\
\bottomrule
\end{tabular}
\inv
\end{table}

\head{R-only is close to R+M because R is not pure refactoring} When \tool{}'s refactoring stage introduces a new model, it also assigns that model an optimal materialization as part of producing the candidate (Section~\ref{sec:materialization}). R-only therefore already runs the materialization tuner on every model it creates, so its gap to R+M in Table~\ref{tab:ablation-rmf} does not measure the value of materialization tuning; it measures only the slice R-only leaves out, the retuning of pre-existing models. 
% \jie{add the sentence below for section 4.4 comments} \linc{I still think putting the numbers in the table would be better. Why not directly put all the percentage numbers in Table 4? Then, you can reference these numbers if needed throughout Section 4.3, instead of just this part.} 
That slice is small: only 16 of Tuva's 255 models have their materialization retuned (Table~\ref{tab:coverage}), and the R-only to R+M gap reflects exactly those 16. M-only trails R-only for the complementary reason: it revises materialization only on the original DAG and cannot create models. Even so, M cannot be dropped in favor of R+F. On Tuva the materialization that R assigns to its new models already sits close to the project-wide optimum, so M's extra retuning adds little; this overlap varies across projects, and where it is smaller M contributes more.

% \head{R-only is close to R+M because R is not pure refactoring} When \tool{}'s refactoring stage introduces a new model, it also assigns that model an optimal materialization as part of producing the candidate (Section \ref{sec:materialization}). \linc{broken reference} \jie{Fixed.} The R-only variant therefore already tunes newly created models; what M adds is project-wide retuning of pre-existing models, a smaller marginal change, so the gap between R-only (49.1\%) and R+M (54.7\%) is modest. M-only (37.0\%) trails R-only because it cannot create models and can only revise materialization on the original DAG. \linc{If M only adds marginally on top of R, shouldn't we just remove M from the pipeline and use R+F?}

% \head{Refactoring and materialization are coupled} R+M (54.7\%) exceeds the larger singleton (R: 49.1\%) by 5.6 points: the refactored DAG admits a different optimal materialization than the original, and the joint pass finds savings neither stage produces in isolation. This makes the Section~\ref{sec:eval-baselines} argument concrete: pattern 4 of Section~\ref{sec:eval-baselines} contributes to R+M but to neither R nor M alone. \linc{Eh this paragraph and the paragraph above actually conflicts with each other... this paragraph says that both stages are important, but the previous paragraph says that what M adds over R is only marginal change. So which is our conclusion? We're referencing exactly the same number here.}

\head{Frequency's increment depends on how much over-scheduled work the prior dimensions leave behind} F alone delivers 11.0\%, adds 17.0\% on top of M (54.0\% vs.\ 37.0\%) and 13.0\% on top of R+M (67.7\% vs.\ 54.7\%). F's reach is bounded by the over-scheduled work present in the project, and that pool depends on the prior dimensions: refactoring can move work onto the right cadence and remove it from F's surface (splitting a fast-cadence model into a slow rollup and a fast lookup), while materialization tuning leaves the over-schedule pool largely intact, so F has more to remove on top of M than on top of R+M. F's increment never collapses to zero, which shows it is orthogonal to R and M rather than a redundant pass; and because that increment varies with the prior dimensions, the three interact rather than add independently.

% \head{Frequency's increment depends on what came before it} \linc{it's unclear what does ``what'' means here? Models? Improvements from other optimizations? We should be explicit.} F alone delivers 11.0\%, adds 17.0 points on top of M ($54.0$ vs.\ $37.0$) and 13.0 on top of R+M ($67.7$ vs.\ $54.7$). F's reach is bounded by the over-scheduled work present in the project, and that pool depends on the prior dimensions: refactoring can move work onto the right cadence and remove it from F's surface (splitting a fast-cadence model into a slow rollup and a fast lookup), while materialization tuning leaves the over-schedule pool largely intact, so F has more to remove on top of M than on top of R+M. The increment never collapses, which confirms F is structurally orthogonal to R and M rather than a redundant pass, and its size confirms the three dimensions interact in non-trivial ways. \linc{size of what? time improvements?}

\subsubsection{Dataflow Analysis}
\label{sec:eval-ablation-dataflow}

This experiment isolates the contribution of the dataflow analysis that produces \tool{}'s candidate set (Section~\ref{sec:candidate-identification}): does structural targeting direct the LLM stage to more productive regions of the DAG than alternatives that do not use it? We hold the LLM-call budget fixed across every configuration we compare, so the comparison reflects how well a fixed budget is spent rather than how many invocations are made. In the standard pipeline, dataflow analysis selects at most 20 subgraphs on Tuva, each capped at 60 nodes with 3 diagnostic-and-correction attempts; we hold this budget fixed across two alternatives. Table~\ref{tab:ablation-dataflow} reports the result.

\begin{table}[tb]
\centering
\caption{Dataflow Analysis: targeting ablation on Tuva (\% slot reduction over the original project) under a fixed LLM-call budget across all arms.}
\inv
\label{tab:ablation-dataflow}
\small
\begin{tabular}{lcc}
\toprule
Selection rule & Single-run slot & Freq-weighted slot \\
\midrule
Whole-DAG                  & \multicolumn{2}{c}{no valid DAG} \\
Random subgraphs           & $43.1$ & $58.0$ \\
Dataflow-guided (\tool{})  & $\mathbf{51.1}$ & $\mathbf{67.7}$ \\
\bottomrule
\end{tabular}
\inv
\end{table}

\head{Whole-DAG refactoring} The first alternative hands the entire project to the LLM under the same 3-attempt budget, testing whether bounding its attention to a region is needed at all. On Tuva the whole DAG fits in context (134k input tokens) and analysis succeeds, surfacing most of the opportunities our pipeline finds, but implementation never produces a runnable project. The response must restate every rewritten model in full, so it grows long enough to truncate mid-output and leave dangling references; cross-model references grow inconsistent as a fix to one model breaks an assumption in another; and because every broad attempt breaks somewhere, the correction loop retreats in scope across the three attempts (51, then 35, then 8 models). A broad rewrite captures most of the gain but touches too many interdependent models to verify, while one narrow enough to verify captures only a fraction. Dataflow analysis resolves this by making each subgraph small enough to verify while still reaching the high-value regions one at a time.

\head{Random subgraphs} The second alternative changes only the selection rule, holding the subgraph count, size cap, and refactoring procedure fixed: we draw a model uniformly at random, take its parents, children, and siblings as a subgraph, and repeat until the count matches the standard pipeline, discarding any neighborhood over the node cap. This isolates the contribution of \emph{where} the pipeline chooses to look. As Table~\ref{tab:ablation-dataflow} shows, random selection reaches 43.1\% single-run and 58.0\% frequency-weighted slot reduction, against 51.1\% and 67.7\% for dataflow-guided selection. The gap is a structural mismatch: an edge-based subgraph groups a model with its dependency neighbors, but the models worth refactoring together are elsewhere in the project, computing the same pattern over different inputs without sharing a single edge, so no edge-based neighborhood contains both and widening it only adds irrelevant dependency-adjacent models. Dataflow-guided selection groups on the structure of the computation rather than on dependency edges, placing the models that compute the same pattern in one subgraph regardless of where they sit in the DAG.

\subsection{System Characterization}
\label{sec:eval-system}

This subsection asks whether the machinery behind the gains is sound and affordable: the equivalence-verification loop that catches unsafe rewrites, the LLM cost of running \tool{} end to end, and the accuracy of the learned cost model that drives materialization.

\subsubsection{Equivalence-Verification Effectiveness}
\label{sec:eval-verification}

The question for the verification loop (\S\ref{sec:llm-refactoring}) is whether it catches unsafe rewrites without discarding safe ones, under a budget of three correction attempts per subgraph. Of the 20 candidate subgraphs on Tuva, 12 fit within the size cap and reach verification (Table~\ref{tab:verification}). Only a quarter pass on the first attempt, and the rewrites the diagnostics reject fall into two recurring patterns a per-query checker cannot detect. The dominant one is a grain collapse: the rewrite reads a value from a pre-deduplicated sibling that keeps one row per key, silently dropping the secondary rows the original aggregates over (Figure~\ref{fig:critic-catch}). The second is algorithm substitution: a cheaper algorithm whose output differs from the original, such as a window-based interval merge that yields different groupings than the pairwise merge it replaces. Yet none of the 12 exhausts the budget: every rejected subgraph passes within a second attempt once its violated invariants are returned to the rewriter, so the loop is a gate rather than a dead end. On Stripe most candidates pass on the first attempt, since its smaller DAG and simpler SQL leave the rewrites less error-prone.

\begin{table}[tb]
\centering
\caption{Equivalence-verification outcomes on Tuva's 12 refactoring subgraphs (three-attempt budget each).}
\inv
\label{tab:verification}
\small
\begin{tabular}{lr}
\toprule
Outcome & Subgraphs \\
\midrule
Pass within 1 attempt        & 25\% (3/12) \\
Pass within 2 attempts       & 100\% (12/12) \\
Never pass (budget exhausted) & 0\% \\
\bottomrule
\end{tabular}
\end{table}

\subsubsection{LLM Cost}
\label{sec:eval-llm-cost}

Running \tool{} is a one-time, design-time cost, paid once per project and amortized over every scheduled run of the optimized pipeline. Optimizing Tuva's 375 models with GPT-5.4 cost \$11.51 and 2.4M tokens, about \$0.03 per model, dominated by the rewriting stage (Table~\ref{tab:llmcost}). This is small beside the recurring warehouse cost the optimized pipeline removes on every run.

\begin{table}[tb]
\centering
\caption{End-to-end LLM cost of optimizing Tuva (GPT-5.4)}
\inv
\label{tab:llmcost}
\small
\begin{tabular}{lccccc}
\toprule
 & Proposal & Filtering & Rewriting & Verification & Total \\
\midrule
Cost (\$)  & 2.00 & 0.96 & 5.68 & 2.88 & 11.51 \\
Share (\%) & 17   & 8    & 49   & 25   & 100 \\
\bottomrule
\end{tabular}
\inv
\end{table}

\subsubsection{Cost Model Accuracy}
\label{sec:eval-cost-model}

The learned cost model only needs to rank refactoring and materialization options within a project, not predict absolute runtimes or generalize across projects. Fit per project on the 156 instrumented Tuva models, its slot-time predictor, the quantity the ILP minimizes, reaches R$^2$=0.74 from static SQL and DAG features alone. Cardinality is harder (R$^2$=0.28), since row counts depend on data content that static features cannot see, but directional estimates are enough for ranking.

%% file: sections/5_related.tex
% !TEX root = ../VLDB-2026-DAGSmith-Dependency-Aware-Rewriting-for-dbt-Style-SQL-Pipelines.tex

% !TEX root = ../VLDB-2026-DAGSmith-Dependency-Aware-Rewriting-for-dbt-Style-SQL-Pipelines.tex
\section{Related Work}
\label{sec:related}

\head{Positioning}
\tool is, to the best of our knowledge, the first system to treat the explicit pipeline DAG as the unit of dependency-aware rewriting, restructuring node boundaries and deciding refactoring, materialization, and refresh frequency jointly. The closest lines of work each address only part of this in isolation.

\head{Single-query SQL rewriting}
Rule-based rewriters~\cite{wetune,learnedrewrite}, program-synthesis rewriters~\cite{slabcity}, and LLM-based rewriters~\cite{genrewrite,rbot,llmr2} transform a query into an equivalent form before the database optimizer plans it; two of our baselines, LearnedRewrite~\cite{learnedrewrite} and GenRewrite~\cite{genrewrite}, are of this kind. All optimize one statement in isolation and leave the surrounding dependency graph unchanged.

\head{Multi-query optimization and view selection}
Multi-query optimization~\cite{sellis1988multiple,roy2000efficient} and common-subexpression sharing~\cite{scopeCSE,similarSubexpr} reuse intermediate results across statements, and materialized-view selection~\cite{agrawal2000automated,harinarayan1996implementing} precomputes chosen results and rewrites consumers to read them; our cross-model CSE baseline follows this line~\cite{scopeCSE}. These techniques match on syntactic structure and select persistence among results that already exist, without recognizing the same computation written in dissimilar SQL or trimming results no downstream output needs.

% They do not recognize the same logical computation written in dissimilar SQL, as in Figure~\ref{fig:strategy-c-sql-reuse},  and they keep every output intact, sharing how results are computed but never which are needed \linc{$\leftarrow$ I don't understand this last phrase...}; \tool reads downstream demand to trim redundant computation and simplify the dependency graph (Figure~\ref{fig:strategy-c-sql-edge}).

\head{View maintenance and refresh scheduling}
Incremental view maintenance~\cite{ivm} and streaming materialized-view systems~\cite{materialize,risingwave} keep derived results fresh by processing only the rows that change, lowering the cost of each refresh. \tool instead aligns how often each node runs with how often its inputs change and its outputs are consumed, a complementary direction that could compose with theirs.

\head{Workflow orchestration}
dbt~\cite{dbtWhatExactly}, SQLMesh~\cite{sqlmeshOverview}, Airflow~\cite{airflowDags}, Dagster~\cite{dagsterAssets}, Prefect~\cite{prefectTasks}, Argo~\cite{argoDag}, Luigi~\cite{luigiDocs}, and Databricks Workflows~\cite{databricksJobs} express recurring transformations and their dependencies as a DAG, the setting \tool targets. Their schedulers optimize execution order, invalidation, retries, and resource allocation, but leave the SQL semantics of each node untouched, which is where \tool's reductions originate.

%% file: sections/6_conclusion.tex
% !TEX root = ../VLDB-2026-DAGSmith-Dependency-Aware-Rewriting-for-dbt-Style-SQL-Pipelines.tex
\section{Conclusion}
\label{sec:conclusion}

Explicit SQL pipeline DAGs create an optimization setting that is richer than traditional one-query-at-a-time rewriting. Once dependencies, downstream demand, final outputs, materialization choices, and refresh behavior are visible, the optimizer can restructure the pipeline itself rather than merely improve individual statements. \tool is a step toward that goal: it treats dependency-aware pipeline optimization as a joint problem over graph structure, materialization, and frequency, and combines structural analysis with LLM-guided refactoring and verification to explore that space.

% The core claim is that recurring SQL pipeline DAGs expose whole-pipeline optimization opportunities that are invisible to query-local rewriting, and exploiting them requires a system designed for dependency-aware reasoning.
% \linc{I actually think we can remove this last paragraph if we're short on space. It's a bit redundant.}